\documentclass[aps, 10pt, prd,   
amsmath,amssymb,nofootinbib,superscriptaddress,hyperref,nobalancelastpage,
onecolumn,notitlepage,biblatex]{revtex4-1} 
\usepackage{slashed}
\usepackage{booktabs}
\usepackage{braket}
\usepackage{float}
\usepackage[utf8]{inputenc}
\usepackage{graphicx}
\usepackage{subfigure}
\usepackage{latexsym}
\usepackage{epic}
\usepackage{eepic}
\usepackage{cancel}
\usepackage{epsfig}
\usepackage{color}
\usepackage{natbib}
\usepackage{hyperref}
\usepackage{array}
\usepackage{bm}
\usepackage{multirow}

\hypersetup{
	colorlinks=true,
	linkcolor=myviolet,
	urlcolor=myviolet,
	citecolor=myviolet
}
\usepackage[normalem]{ulem}

\usepackage{longtable}
\usepackage{pbox}
\usepackage{placeins}

\definecolor{darkred}{rgb}{0.0,0,0.6}
\definecolor{verdec}{cmyk}{0.98,0,0.59,0.15}
\definecolor{verdes}{cmyk}{0.92,0,0.59,0.4}
\definecolor{mypink2}{RGB}{219, 48, 122}
\definecolor{myviolet}{RGB}{205,105,255}

\newcommand{\AddrHBNI}{
	Homi Bhabha National Institute, BARC Training School Complex, Anushakti Nagar, Mumbai 400094, India }
\newcommand{\AddrIOP}{
	Institute of Physics, Sachivalaya Marg, Bhubaneswar 751005, India}
\renewcommand\thesection{\arabic{section}}
\renewcommand\thesubsection{\arabic{section}.\arabic{subsection}}

\begin{document}
\title{\boldmath An Alternate Left-Right Symmetric Model with Dirac Neutrinos}
	 
\author{Siddharth P. Maharathy}\email{siddharth.m@iopb.res.in}
\affiliation{\AddrIOP}
\affiliation{\AddrHBNI}
	
\author{Manimala Mitra}\email{manimala@iopb.res.in}
\affiliation{\AddrIOP}
\affiliation{\AddrHBNI}

\author{Agnivo Sarkar}\email{agnivo.sarkar@iopb.res.in}
\affiliation{\AddrIOP}
\affiliation{\AddrHBNI}

\begin{abstract}
	We study a different variant of Left-Right Symmetric Model, incorporating Dirac type neutrinos. In the absence of the bi-doublet scalars, the possibility of a universal seesaw type of mass generation mechanism for all the Standard Model charged fermions have been discussed. The model has been constructed by extending the Standard Model particle spectrum with heavy vector-like fermions as well as different scalar multiplets. We have shown that this model can generate non zero neutrino mass through loop mediated processes. The parameters which are involved in neutrino mass generation mechanism can satisfy the neutrino oscillation data for both normal and inverted hierarchy. The lightest charged Higgs plays a crucial role in neutrino mass generation mechanism and can have mass of $\mathcal{O}[\text{GeV}]$. We have systematically studied different constraints which are relevant for the charged Higgs phenomenology. In addition to that we also briefly discuss discovery prospects of the charged Higgs at different colliders. 
\end{abstract}

\maketitle
\section{\label{Introduction}Introduction} 

The Standard Model (SM) of particle physics consistently describes the dynamics of three out of four fundamental forces of nature. The last missing puzzle of the theory, the Higgs boson which arises due to the manifestation of the Brout-Englert-Higgs~(BEH) mechanism \cite{Englert:1964et,Higgs:1964ia,Higgs:1964pj,Guralnik:1964eu,Higgs:1966ev,Kibble:1967sv} was discovered by the ATLAS \cite{ATLAS:2012yve} and CMS \cite{CMS:2012qbp} collaboration in the beginning of last decade. Since its discovery the experimentalists have studied its properties with a great detail and the information accumulated from these analysis solidify the triumph of the SM. Having said that, the questions that revolve around the SM from both the theoretical as well as experimental stand point motivate physicists to consider the SM as a minimal theory and an extended version of this model is required to address these questions. 

Out of the various shortcomings of the SM,  eV light neutrino mass and their mixing remains one of  the most significant one. In SM  these neutral leptons are considered to be massless, which is in disagreement with the measured neutrino oscillation data \cite{deSalas:2017kay}. The simplest solution to explain light neutrino masses is to add three right handed SM gauge singlet neutrino fields $\nu_R$ with the existing SM particle contents and generate  Dirac mass term. The  eV scale masses of the neutrinos however force to choose   tiny Yukawa coupling  $Y_{\nu} \sim 10^{-11}$. The presence of  small $Y_{\nu}$ introduces additional Hierarchy problem to the theory. As neutrino is a neutral fermion, another compelling solution  is to consider it as a Majorana particle instead of Dirac particle. With this, small neutrino masses can be generated from the $d=5$ lepton number violating~(LNV) Weinberg operator~\cite{Weinberg:1979sa} via seesaw mechanism \cite{Minkowski:1977sc, Cheng:1980qt,Schechter:1980gr, Foot:1988aq}. 

If neutrinos are Majorana particles, they give rise to different lepton number violating (LNV) signatures, such as,  neutrino-less double beta decay, LNV decays of mesons, as well as, LNV processes at the LHC.  The null results from these experiments   enforce stringent bounds on corresponding processes, such as, on the half-life of $0\nu \beta \beta$ process in  case of neutrino-less double beta decay, on the relevant production cross-section in case of LHC etc. As a result the Dirac type neutrino mass models also remain a much motivated option, see \cite{Roy:1983be,Saad:2019bqf,Ma:2014qra,Ma:2015mjd,Bonilla:2016zef,CentellesChulia:2016rms,Bonilla:2016diq,Ma:2016mwh,Wang:2016lve,Borah:2017dmk,Yao:2018ekp,Reig:2018mdk} for few of the previous studies. To avoid any additional hierarchy problem,  tiny neutrino masses can be generated via  loop diagrams where relatively heavy particles flow in the  loop and shift the neutrino mass pole to a non zero value. The mass generation in this scheme does not demand  unnatural fine-tuning of any model parameter.  One may recall  examples such as,  Scotogenic Model \cite{Ma:2006km}, Zee-Babu Model \cite{Zee:1985id,Babu:1988ki,Nebot:2007bc} \emph{etc}. The work by Ma and Sarkar \cite{Ma:2017kgb} has systematically summarised the different avenues for the Dirac neutrinos to achieve masses via radiative process. In this work, we consider an alternate version of  gauge extended Left-Right Symmetric Model (LRSM)  with two scalar doublets and a charged scalar singlet and explore Dirac neutrino mass generation in detail. In LRSM, one can naturally introduce  right-handed neutrino fields $\nu_R$ which belong to  doublets under $SU(2)_{R}$ gauge group. In the absence of a scalar bi-doublet which is charged under both the $SU(2)_{L/R}$ groups, the fermions in the model remain massless. One can generate tree level masses for quarks and charged leptons  by introducing heavy vector-like fermions. The mass generations  scheme is  known as universal seesaw mechanism \cite{Davidson:1987mh,Mohapatra:1987nx,Ma:1989tz,Ma:2017kgb,Brahmachari:2003wv}.  In absence of any gauge singlet fermionic field, the neutrino however acquires mass via radiative processes. The idea to generate  one loop diagrams, relevant for neutrino mass generation is straightforward. One needs to connect $\nu_{L}$ and $\nu_{R}$ with an internal fermion and scalar line. With this set up one can come up with four independent structures out of which two  correspond to Zee-Babu and Scotogenic model. The remaining two possibilities were  first discussed in   \cite{Ma:2017kgb} with  exotic scalar representations  playing  crucial role in  neutrino mass generation. 

In this paper we present a detail description of a variant of LRSM model where the left right gauge sector does not mix through a scalar bi-doublet. Using the idea of universal seesaw mechanism we generate the masses for the SM-like charged fermions with the help of heavy vector-like fermions. As a result, the model possesses a natural explanation for the observed fermion mass hierarchy between the SM quarks and leptons. On the other hand the Dirac-like neutrinos in this model remain massless at tree level. Apart from the extended gauge and fermion sector the model also contains an enlarged scalar sector with exotic scalar representations. These scalar fields generate the tiny mass of neutrinos through loop-mediated process. In this model the mechanism for the neutrino mass generation can be understood as the amalgamation of two recently proposed method, mentioned in \cite{Ma:2017kgb}. In addition to the neutrino mass generation, we also discuss the direct and indirect constraints on the model parameters coming from ATLAS di-lepton+MET search, LEP mono-photon search as well as from Big-Bang Nucleosynthesis. We analyse the production cross-section as well as branching ratios of the gauge singlet charged Higgs and briefly discuss discovery prospect at different colliders.

The article is organized in the following manner. In Sec.\,\ref{Sec:Model}, we begin with the general set up of the model and describe the gauge, scalar and fermion sector in a detailed manner. In  Sec.\,\ref{Sec:MassNu}, we discuss   the mechanism of neutrino mass generation and show the allowed region of the parameter space which satisfy the well-measured neutrino oscillation data. In Sec.\,\ref{Sec:Pheno}, we discuss the phenomenological aspects of our model focusing on  the lightest singly charged scalar present in our model. By scrutinizing all the possible direct as well as indirect searches, we  show the available parameter space.  Finally, in Sec.\,\ref{Sec:conclu}, we summarize our findings.

\section{\label{Sec:Model}A Brief Review of the Model}

\subsection{General Model Setup}
\label{SubSec:Setup}
\noindent 
 
 The model we have considered here is based on the gauge group: $SU(3)_{C}\times SU(2)_{L}\times SU(2)_{R}\times U(1)_{B - L}$, which is a minimal extension of the SM gauge symmetry with an additional $SU(2)_R$ gauge symmetry. We further assume that the $SU(2)_{L}$ group resembles with the SM counterpart. The gauge coupling corresponding to each gauge groups are denoted as $g_{L}$, $g_{R}$ and $g_{B - L}$ respectively. Two scalar doublet fileds $\Phi_{L}$ and $\Phi_{R}$ which are charged under $SU(2)_{L}$ and $SU(2)_{R}$ respectively develop \emph{vev}s and invoke the spontaneous symmetry breaking in this model. The $\Phi_{R}$ field breaks the electroweak symmetry down to $SU(2)_{L}\times U(1)_{Y}$ at the energy scale $v_{R}$. After that the other scalar doublet $\Phi_{L}$ develops a \emph{vev} ($v_{L}$) and breaks the residual symmetry down to $U(1)_{EM}$. Based upon the electroweak symmetry breaking (EWSB) pattern, the charge operator in this model can be defined as \cite{Marshak:1979fm}, $\hat{Q} = \frac{T_{3L}}{2} + \frac{T_{3R}}{2} + \frac{Y_{B - L}}{2}$\footnote{The subscript in each of the generators denotes the corresponding gauge group.}.  In Eq.\,\ref{Eq:PhiLPhiR} we present the explicit form of these scalar doublet fields. 
\begin{equation}
		\Phi_L = 
		\begin{bmatrix}
		\phi_L^+ \\ \phi_L^0	
		\end{bmatrix} \Rightarrow \begin{bmatrix}
		\phi_L^+\\ \frac{v_L}{\sqrt{2}}+\frac{h_L^0 + i \pi_L^0}{\sqrt{2}}
		\end{bmatrix}
, ~~~~ 
\Phi_R = 
		\begin{bmatrix}
		\phi_R^+ \\ \phi_R^0	
		\end{bmatrix} \Rightarrow \begin{bmatrix}
		\phi_R^+\\ \frac{v_R}{\sqrt{2}}+\frac{h_R^0 + i \pi_R^0}{\sqrt{2}}
		\end{bmatrix}  
\label{Eq:PhiLPhiR}
\end{equation}
\noindent
In our notation, $\phi^{+}_{L}$ and $\phi^{+}_{R}$ denotes the charged goldstone fields whereas the $\pi^{0}_{L}$ and $\pi^{0}_{R}$ denotes the neutral goldstone bosons. The $h^{0}_{L}$ and $h^{0}_{R}$ are the CP-even Higgs bosons which are written in the gauge basis. The $v_{L}$ and $v_{R}$ are the \emph{vev'}s correspond to the doublet fields $\Phi_{L}$ and $\Phi_{R}$ respectively. Throughout this paper we fix the value of $v_L$ to be EWSB scale $v = v_L \simeq$ 246 GeV.   
Apart from the scalar doublets, the model contains other exotic scalar fields which play a crucial role for the neutrino mass generation in this model. It is important to note that these exotic fields do not take part in the EWSB. In Eq.\,\ref{Eq:zetachi} we present the explicit form of these fields. The $\zeta_{L}$/$\zeta_{R}$ are $SU(2)_L$/$SU(2)_R$ doublets which contain both the doubly and singly charged components whereas the $\chi^{\pm}$ is a $SU(2)_{L/R}$ singlet. In Table.\,\ref{Tab:Scalar} we present the gauge charges of these fields with respect to the underlying symmetry.
\begin{equation}
{\zeta_L} = 
		\begin{bmatrix}
			\zeta^{++}_{L}\\\zeta^{+}_{L}
		\end{bmatrix}
,~~~~
{\zeta_R} = 
\begin{bmatrix}
	\zeta^{++}_{R}\\\zeta^{+}_{R}
\end{bmatrix}
,~~~~
\chi^{\pm}
\label{Eq:zetachi}
\end{equation}
\noindent
The interesting aspect of our model is the absence of bi-doublet scalar field which is charged under both the $SU(2)_{L}$ and $SU(2)_{R}$ groups. As a consequence of this, the mixing between light and heavy gauge bosons is minimal contrary to the conventional Left-Right symmetric models \cite{Mohapatra:1974gc,Senjanovic:1975rk,Deshpande:1990ip}. We will elaborate on this in the coming section.

\FloatBarrier
	 \begin{table}[H]
	 	\centering
	 	\resizebox{0.5\textwidth}{!}{
		\begin{tabular}{|c|c|}
			\hline
			Multiplets & $SU(3)_C\times SU(2)_R\times SU(2)_L\times U(1)_{B-L}$ \\

\hline
		$ {\Phi_L} $		&

					$     (1,1,2,1)$ \\
	${\Phi_R}$&				$ (1,2,1,1)$  
					\\
	${\zeta_L}$ &				$ (1,1,2,3)$ 
					\\
		${\zeta_R}$ &			$ (1,2,1,3)$ 
					\\
		$\chi^+$ &			$(1,1,1,2)$
					\\
				\hline		
		\end{tabular}
	    }
    \caption{ 
Scalar representation of the model.
    }\label{Tab:Scalar}
\end{table}

The LRSM with an exact parity symmetry has been extensively studied in \cite{Pati:1974yy,Mohapatra:1974gc,Senjanovic:1975rk,Mohapatra:1979ia,Mohapatra:1980qe,Mohapatra:1980yp,Deshpande:1990ip}, where the parity is spontaneously broken at a high energy scale. Another class of LRSMs exits where a discrete left-right symmetry named as D-parity is broken at a higher scale compared to the breaking scale of $SU(2)_R$ gauge group \cite{Chang:1983fu,Chang:1984uy,Sarkar:2004hc,Sahu:2006pf,Gong:2007yv,Borah:2010zq,Patra:2009wc,Deppisch:2014zta}. In this work, we also assume that the parity is broken at a high energy scale and leaves an approximate parity invariance rather than an exact one. We will not discuss the details of parity breaking mechanism because our work focus in an energy regime which is much lower than the scale at parity breaking occurs.

\subsection{\label{SubSec:Gauge}Gauge Sector}
The kinetic energy term corresponds to electroweak gauge sector takes the following form 
\begin{equation}
\mathcal{L}_{K.E.} = - \frac{1}{4}F^{a\mu\nu}_{R}F^{a}_{R\mu\nu} - \frac{1}{4}F^{a\mu\nu}_{L}F^{a}_{L\mu\nu} - \frac{1}{4}B_{\mu\nu}B^{\mu\nu} 
\label{Eq:VKE}
\end{equation}
\noindent
where $B_{\mu\nu} = \partial_{\mu}B_{\nu} - \partial_{\nu}B_{\mu}$ is the field-strength tensor corresponding to $U(1)_{B - L}$ gauge group and $F^{a\mu\nu}_{L/R}$ ($a$ = 1 to 3) is representing the field-strength tensor correspond to $SU(2)_{L/R}$ gauge groups. Out of these seven gauge fields, six become massive after EWSB. To calculate the mass of each of the gauge bosons, one needs to write down the scalar field kinetic energy terms. In Eq.\, \ref{Eq:LGauge} we present the explicit form for this.    
\begin{equation}
\mathcal{L}_{\text{Gauge}} = \left(D^{L}_{\mu}\Phi_{L}\right)^{\dagger}D^{L}_{\mu}\Phi_{L} + \left(D^{R}_{\mu}\Phi_{R}\right)^{\dagger}D^{R}_{\mu}\Phi_{R} + \left(D^{L}_{\mu}\zeta_{L}\right)^{\dagger}D^{L}_{\mu}\zeta_{L} + \left(D^{R}_{\mu}\zeta_{R}\right)^{\dagger}D^{R}_{\mu}\zeta_{R} + \left(D^{S}_{\mu}\chi\right)^{\dagger}D^{S}_{\mu}\chi
\label{Eq:LGauge}
\end{equation}
\noindent
The covariant derivatives corresponds to individual scalar multiplets are defined as - 
\begin{eqnarray}
D^{L}_{\mu} &=& \partial_{\mu} - \frac{ig_{L}}{2}\sigma^{a}W^{a}_{\mu L} - \frac{ig_{B - L}}{2}Y_{B - L}B_{\mu}, \nonumber \\
D^{R}_{\mu} &=& \partial_{\mu} - \frac{ig_{R}}{2}\sigma^{a}W^{a}_{\mu R} - \frac{ig_{B - L}}{2}Y_{B - L}B_{\mu},  \\
D^{S}_{\mu} &=&  \partial_{\mu} - \frac{ig_{B - L}}{2}Y_{B - L}B_{\mu}. \nonumber
\label{Eq:Covariant}
\end{eqnarray}
\noindent
Here $\sigma^{i}$ with i$ \in $[1,3] corresponds to three Pauli matrices.
After EWSB the charged and neutral gauge bosons mass eigen states are defined as \{$W^{\pm}_{L\mu},W^{\pm}_{R\mu}$\}, \{$Z_{L\mu} , Z_{R\mu}$\} respectively.
The mass squared terms for each of the charged gauge bosons $W^{\pm}_{L\mu}$ and $W^{\pm}_{R\mu}$ are,  

\begin{equation}
M^{2}_{W^{\pm}_{L}} = \frac{g^{2}_{L}v^{2}_{L}}{4}, ~~~~~~~~~~ M^{2}_{W^{\pm}_{R}} = \frac{g^{2}_{R}v^{2}_{R}}{4}.
\label{Eq:VCharged}
\end{equation}
\noindent
The absence of cross term between the $W^{\pm}_{L\mu}$ and $W^{\pm}_{R\mu}$ would make them orthogonal to each other even in the gauge basis. This is the consequence of not having any bi-doublet scalar in our model. We identify $W^{\pm}_{L\mu}$ as the SM like charged gauge bosons for our subsequent discussion. The neutral gauge boson mass matrix takes the following form in the gauge basis,  $\{ B_{\mu} $, $W^{3}_{\mu L} $, $ W^{3}_{\mu R} \}$. 
\begin{equation}
M^{2}_{\text{NGB}} = \frac{1}{4}\begin{bmatrix}
g^{2}_{B - L}\left(v^{2}_{L} + v^{2}_{R}\right) & - g_{B - L}g_{L}v^{2}_{L} & - g_{B - L}g_{R}v^{2}_{R} \\
- g_{B - L}g_{L}v^{2}_{L} & g^{2}_{L}v^{2}_{L} & 0 \\
- g_{B - L}g_{R}v^{2}_{R} & 0 & g^{2}_{R}v^{2}_{R}
\end{bmatrix}
\label{Eq:NGaugeM}
\end{equation}
\noindent
The $M^{2}_{\text{NGB}}$ mass matrix has two non-zero and one zero eigenvalues.
The eigenstate corresponds to the zero eigenvalue of the $M^{2}_{\text{NGB}}$ matrix is identified as photon ($A_{\mu}$). On the other hand the neutral gauge bosons $Z_{\mu L} $ and $ Z_{\mu R}$ have non-zero mass eigen values,  $M_{Z_L}$ and $M_{Z_R}$ respectively. In Eq.\,\ref{Eq:neutral_gaugemass}, we present the explicit form of these masses. Out of these two neutral gauge bosons $Z_L$ is the lighter one and $Z_R$ is its heavier counter part. Here after we consider the lighter state ($Z_L$) as the SM $Z$-boson.

 \begin{equation}
	M^{2}_{Z_L} = \frac{1}{4}\frac{v^{2}_{L}g^{2}_{L}}{c^2_{\theta_{W}}}, ~~~~~~ M^{2}_{Z_R} = \frac{1}{4}\left(g^2_{B-L}s^2_{\phi_{W}} + \frac{g_R^2 v_R^2}{c^{2}_{\phi_{W}}}\right). 
	\label{Eq:neutral_gaugemass}
\end{equation}
\noindent
To diagonalise the neutral gauge boson mass matrix one needs to perform three consequtive similarity rotations as discussed in \cite{Frank:2019nid,Ashry:2013loa}. The relation between the gauge basis $\{ B_{\mu} , W^{3}_{\mu L} ,  W^{3}_{\mu R} \}$ and the mass basis $\{A_{\mu},Z_{\mu L},Z_{\mu R}\}$ is defined in Eq.\,\ref{Eq:nuetralgaugemixing}. 

\begin{equation}	
\begin{bmatrix}
B_\mu \\ W^{3}_{\mu L} \\W^{3}_{\mu R}  
\end{bmatrix} = 
\begin{bmatrix}
	c_{\phi_{W}} & 0 & -s_{\phi_{W}} \\ 0 & 1 & 0 \\  s_{\phi_{W}} & 0 & c_{\phi_{W}}
\end{bmatrix}
\begin{bmatrix}
	c_{\theta_{W}} & -s_{\theta_{W}} & 0 \\ s_{\theta_{W}} & c_{\theta_{W}} & 0 \\ 0 & 0 & 1 
\end{bmatrix} 
\begin{bmatrix}
	1 & 0 & 0 \\ 0 & c_{\rho_{W}} & -s_{\rho_{W}}  \\ 0 & s_{\rho_{W}} & c_{\rho_{W}}
\end{bmatrix} 
\begin{bmatrix}
	A_\mu \\ Z_{\mu L} \\ Z_{\mu R}  
\end{bmatrix} 
\label{Eq:nuetralgaugemixing}		
\end{equation}
\noindent
Here $s_i$ and $c_i$ denote sine and cosine of the angle $i$ (where $i = \phi_{W}, \theta_{W}, \rho_{W}$ are the mixing angles) respectively. The angle $\theta_{W}$  represents  the usual \emph{Weinberg angle}. In Eq.\,\ref{Eq:mixing_angle}, we parameterize these mixing angles in terms of various gauge couplings present in the model \cite{Frank:2019nid,Ashry:2013loa}. 
\begin{eqnarray}
	s_{\theta_W} &=& \frac{g_Y}{\sqrt{g_L^2 + g_Y^2} } = \frac{e}{g_L}, ~~~~ s_{\phi_{W}} = \frac{g_{B-L}}{\sqrt{g_{B-L}^2 + g_R^2} } = \frac{g_Y}{g_R} \nonumber \\ 
	t_{2\rho_{W}} &=& \frac{2 c_{\phi_W} c_{\theta_W}g_L g_R s^{2}_{\phi_W} v_L^2}{(g^2_L - c^{2}_{\theta_{W}}g^{2}_{B-L}s^{2}_{\phi_{W}})c^{2}_{\phi_{W}} v_L^2 - c^{2}_{\theta_{W}} g_R^2 v_R^2}
	\label{Eq:mixing_angle}
\end{eqnarray}
\noindent
In Eq.\,\ref{Eq:mixing_angle} we denote $ t_{2\rho_{W}} $ as $\tan(2\rho_{W})$. In addition the $g_Y$ and $e$ stands for the usual hypercharge and electromagnetic couplings respectively. Using normalization condition of the photon state $A_\mu$ one can establish the following relation between the gauge couplings and electromagnetic constant $e$.

\begin{equation}
	\frac{1}{e^{2}} = \frac{1}{g^{2}_{L}} + \frac{1}{g^{2}_{R}} + \frac{1}{g^{2}_{B - L}}~~,~~~~~~~\frac{1}{g_Y^{2}} = \frac{1}{g^{2}_{R}} + \frac{1}{g^{2}_{B - L}} 
\end{equation}

\subsection{\label{SubSec:Scalar}Scalar Sector}

We now turn our attention to the scalar sector of this model. In Eq.\,\ref{Eq:ScalarV} we present the most general potential which is invariant under the $SU(2)_{L}\times SU(2)_{R}\times U(1)_{B - L}$ gauge group. 
\begin{eqnarray}
	\text{V}\left(\Phi_{i}, \chi, \zeta_{j}\right) &=& \lambda_{1}|\Phi_{L}^{\dagger}\Phi_{L}|^{2}  +\lambda_{2}|\Phi_{R}^{\dagger}\Phi_{R}|^{2} + \lambda_{3}\left[\Phi_{L}^{\dagger}\Phi_{L}  \Phi_{R}^{\dagger}\Phi_{R}\right]   + \lambda_{4}|\chi^\dagger\chi|^{2}  \nonumber \\
	&+& \mu_{\chi}^{2}\chi^{\dagger}\chi + \mu_{\zeta}^{2}\left(\zeta_{L}^{\dagger}\zeta_{L} +  \zeta_{R}^{\dagger}\zeta_{R}\right)+ \lambda_{5}\left[\zeta_{L}^{\dagger}\zeta_{L} + \zeta_{R}^{\dagger}\zeta_{R}\right]^{2}  \nonumber \\
	&+&  \lambda_{6}\left(\zeta_{L}^{\dagger}\zeta_{L}   \Phi_{L}^{\dagger}\Phi_{L}   + \zeta_{R}^{\dagger}\zeta_{R} \Phi_{R}^{\dagger}\Phi_{R} \right) + \lambda_{7}\left(\zeta_{L}^{\dagger}\zeta_{L}   \Phi_{R}^{\dagger}\Phi_{R}   + \zeta_{R}^{\dagger}\zeta_{R} \Phi_{L}^{\dagger}\Phi_{L} \right)  \nonumber \\ &+& \lambda_{8}\left[ \Phi_{L}^{\dagger}\Phi_{L} + \Phi_{R}^{\dagger}\Phi_{R}\right]\chi^{\dagger}\chi
	+ \lambda_{9}\left[\zeta_{L}^{\dagger}\zeta_{L} + \zeta_{R}^{\dagger}\zeta_{R}\right]\chi^{\dagger}\chi \label{Eq:ScalarV} \\ &+& \lambda_{10} 
	\left[\Phi_{L}^{\dagger}\zeta_{L}\zeta_{R}^{\dagger}\Phi_{R} + \Phi_{R}^{\dagger}\zeta_{R}\zeta_{L}^{\dagger}\Phi_{L}\right] +  \lambda_{11} \left[\Phi_{L}^{\dagger}\zeta_{L}\zeta_{L}^{\dagger}\Phi_{L} + \Phi_{R}^{\dagger}\zeta_{R}\zeta_{R}^{\dagger}\Phi_{R}\right]  \nonumber \\
	&+& \left[\Lambda^\prime\chi^-\phi_L^\dagger\zeta_L+\Lambda^{\prime\prime}\chi^-\phi_R^\dagger\zeta_R+h.c \right] - \mu^2_L \Phi_{L}^{\dagger}\Phi_{L} - \mu^2_R \Phi_{R}^{\dagger}\Phi_{R} \nonumber
\end{eqnarray}
\noindent
Both the doublet fields, $\Phi_L$ and $\Phi_R$ ,that are charged under  $SU(2)_L$ and $SU(2)_R$ gauge group acquire \emph{vev}s and break the underlying symmetry down to $U(1)_{EM}$. The rest of the scalar multiplets do not acquire any \emph{vev} and do not participate in the EWSB. Using the minimization condition \cite{Mohapatra:2014qva} of the scalar potential $\text{V}\left(\Phi_{i}, \chi, \zeta_{j}\right)$ \emph{i.e.} $\frac{\partial \text{V}}{\partial v_L} = \frac{\partial \text{V}}{\partial v_R} = 0$, one can express different \emph{vev}s in terms of the scalar parameters ($\lambda_{i}$) in the following manner
\begin{eqnarray}
	\frac{v_L^2}{2} = \frac{\lambda_3 \mu_R^2 - 2 \lambda_2\mu_L^2}{\lambda_3^2 - 4 \lambda_1 \lambda_2}
	\label{tadpole_vl}
\end{eqnarray}
\begin{eqnarray}
	\frac{v_R^2}{2} = \frac{\lambda_3 \mu_L^2 - 2 \lambda_1\mu_R^2}{\lambda_3^2 - 4 \lambda_1 \lambda_2}
\label{tadpole_vr}
\end{eqnarray}
\noindent
From Eq.\,\ref{tadpole_vl} and Eq.\,\ref{tadpole_vr} it is evident that in case of $\mu_L = \mu_R$ both the \emph{vevs} ($v_L$ and $v_R$) become equal. As a consequence an exact parity invariance would emerge. A difference between $v_L$ and $v_R$ can however be generated through different radiative corrections. In the beginning of Sub-section.\,\ref{SubSec:Setup}, we have mentioned that we consider parity to be broken at a high energy scale and leaves an approximate parity invariance rather than an exact one. As a result one can assume $\mu_L \neq \mu_R$. The assumption is valid as we are working at a scale where parity invariance is not realized. 
Using this assumption we will fix $v_{R} \gg v_{L}$ for our further study, with the assumption $\mu_L \gg \mu_R$. In absence of the exact parity invariance, one can also write two soft breaking terms which are proportional to $\Lambda^{'}$ and $\Lambda^{''}$ respectively. In our further study, we consider a very small $\Lambda^{'}$ and $\Lambda^{''}$ for which, as we will show in the next section, one of the charged Higgs ($\chi_\pm$) become decoupled. 

\noindent
 Along with the quadratic field terms the scalar potential contains various mixing terms which are permitted by the gauge symmetry of the model. Moreover these mixing terms can bring forth rich phenomenological aspects for this model. The neutral component of $\Phi_L $ and $ \Phi_R $ \emph{i.e.} $ h_L^0  $ and  $ h_R^0 $, mix among each other through the potential term proportional to $\lambda_{3}$. The $2\times2$ mass matrix corresponding to these real scalar fields takes the following form in the gauge basis.
 
\begin{equation}
M^{2}_{\text{Even}} = \frac{1}{2}\begin{bmatrix}
\mathcal{M}_{11} & \mathcal{M}_{12} \\
\mathcal{M}_{12} & \mathcal{M}_{22} 
\end{bmatrix} 
= \frac{1}{2}
\begin{bmatrix}-\mu_L^2+3\lambda_{1}v_L^2+\frac{\lambda_{3}}{2}v_R^2 &\frac{\lambda_{3}}{2}v_R v_L \\ \frac{\lambda_{3}}{2}v_R v_L & -\mu_R^2+3\lambda_{1}v_R^2+\frac{\lambda_{3}}{2}v_L^2
\end{bmatrix}
\label{Eq:CPMass}
\end{equation}
\noindent
The off-diagonal terms of the above mass matrix are equal due to the hermiticity. After diagonalising $M^{2}_{\text{Even}}$, one can obtain the following two neutral scalar mass eigenstates  
\begin{eqnarray}
H^0 & = &\cos\theta h_{L}^0 + \sin\theta h_{R}^0 \nonumber \\
h_{sm} & = &- \sin\theta h_{L}^0 + \cos\theta h_{R}^0
\end{eqnarray}
where the mixing angle $\theta$ is defined as 
\begin{eqnarray}
\tan2\theta & = & \frac{2\mathcal{M}_{12}}{\mathcal{M}_{22} - \mathcal{M}_{11}} 
\end{eqnarray}
\noindent
and $M_{H^0/h_{sm}}$ are the eigenvalues correspond to each these mass states. The explicit form of these mass eigenvalues are
\begin{equation}
M^{2}_{H^0/h_{sm}} = \frac{1}{2}\left[ \mathcal{M}_{11} + \mathcal{M}_{22} \pm \sqrt{\left(\mathcal{M}_{22} - \mathcal{M}_{11}\right)^{2} + 4\mathcal{M}^{2}_{12}} \right]
\end{equation}
 Hereafter in our discussion we denote the $h_{sm}$ as the SM-like Higgs boson which has been discovered at the LHC. In the limit $v_{R} \gg v_{L}$ the mass of the heavy Higgs $H^0$ is proportional to $SU(2)_{R}$ breaking scale $v_{R}$. In this regime the mixing between the light Higgs $h_{sm}$ and the heavy Higgs $H^0$ is minimal and the value of scalar parameter $\lambda_{1}$ will be in the range of SM scalar self-coupling $\lambda_{h}$. \\ 
\noindent  
 Apart form the neutral Higgs bosons the particle spectrum of the model contains both singly and doubly charged scalars, $\zeta^{\pm}_{L},\zeta^{\pm}_{R},\chi^\pm, \zeta^{\pm\pm}_{L} , \zeta^{\pm\pm}_{R} $, after symmetry breaking. The underlying electroweak symmetry does not permit the doubly charged Higgses to mix among each other. In Eq.\,\ref{Eq:doubly_Hc} we write down the explicit mass terms for the doubly charged fields by adding up the appropriate quadratic order field terms.          
\begin{eqnarray}
	M^{2}_{\zeta^{++}_{R}} & = &\mu^{2}_{\zeta} + \frac{\lambda_{6} v^{2}_{R} + \lambda_{7} v^{2}_{L} }{2}\nonumber \\
	M^{2}_{\zeta^{++}_{L}} & = &\mu^{2}_{\zeta} + \frac{\lambda_{6} v^{2}_{L} + \lambda_{7} v^{2}_{R} }{2}
\label{Eq:doubly_Hc}
\end{eqnarray}

\noindent
The absence of an explicit mixing term makes the doubly charged scalars ($ \zeta^{\pm\pm}_{L} , \zeta^{\pm\pm}_{R} $) to be orthogonal even in the gauge basis. In contrast to that a similar conclusion can not be made for the singly charged scalars: $ \chi^\pm,\zeta^{\pm}_{L},\zeta^{\pm}_{R}$.  The terms associated with $\Lambda^{''}$, $\Lambda^{'}$ and $\lambda_{10}$ give rise to nontrivial cross terms between various singly charged scalar fields in this model. In Eq.\,\ref{Eq:Singly_Hc}, we present the explicit form of the singly charged scalar mass matrix in the gauge basis $\{ \chi^{\pm} ,  \zeta_L^{\pm}, \zeta_R^{\pm} \}$. 
\begin{eqnarray}
M_{\pm}^{2} & = &
\begin{bmatrix}
M_{\chi} & M_{\chi L} & M_{\chi R} \\
M_{\chi L} & M_{L} & M_{LR} \\
M_{\chi R} & M_{LR} & M_{R}
\end{bmatrix}
\nonumber \\
 & = &
\begin{bmatrix}
	\mu^{2}_{\chi} + \frac{\lambda_{8}}{2}\left(v^{2}_{L} + v^{2}_{R}\right)  & \frac{\Lambda^{'}v_{L}}{\sqrt{2}} & \frac{\Lambda^{''}v_{R}}{\sqrt{2}} \\
	\frac{\Lambda^{'}v_{L}}{\sqrt{2}} & \mu^{2}_{\zeta} + \frac{\lambda_{6} v^{2}_{R} + \lambda_{7} v^{2}_{L} }{2}+ \frac{\lambda_{11}}{2}v^{2}_{L} & \frac{\lambda_{10}}{2}v_{L}v_{R} \\
	\frac{\Lambda^{''}v_{R}}{\sqrt{2}} & \frac{\lambda_{10}}{2}v_{L}v_{R} & \mu^{2}_{\zeta} + \frac{\lambda_{6} v^{2}_{L} + \lambda_{7} v^{2}_{R}}{2}  + \frac{\lambda_{11}}{2}v^{2}_{R} 
\end{bmatrix}
\label{Eq:Singly_Hc}
\end{eqnarray}
\noindent 
The next step is to diagonalise the above $3\times3$ matrix to derive the mass eigenstates and eigenvalues of the singly charged scalars respectively. In order to do so, one need to construct the appropriate unitary matrix that would diagonalise $M^{2}_{\pm}$ and establish the relation between the gauge basis and mass basis. One can obtain this result by applying the Jacobi prescription for matrix diagonalisation. For simplicity we assume the soft-breaking terms proportional $\Lambda^{'}, \Lambda^{''} \sim 0$, which decouples the singlet charged scalar $(\chi^{\pm})$ from the rest of the two singly charged scalars and its mass is equal to $M_{\chi}$. In this scenario, the above matrix in Eq.\,\ref{Eq:Singly_Hc} takes the following  block diagonal form
\begin{equation}
M_{\pm}^{2}  =
\begin{bmatrix}
M_{\chi} & 0 & 0 \\
0 & M_{L} & M_{LR} \\
0 & M_{LR} & M_{R}
\end{bmatrix}
\label{Eq:Hcblock}
\end{equation}
To diagonalise the above mass matrix one can use the following rotational matrix 
\[ U_{\pm} = \begin{bmatrix}
1 & 0 & 0 \\
0 & \cos\theta_{\pm} & -\sin\theta_{\pm} \\
0 & \sin\theta_{\pm} & \cos\theta_{\pm}
\end{bmatrix} \]
 where the mixing angle $\theta_{\pm}$ can be defined as 
\begin{eqnarray}
\tan{2 \theta_\pm} =  \frac{2M_{LR}^{2}}{M_{R}^{2} - M_{L}^{2}}
\label{Eq:Singly_chrg_mixing}
\end{eqnarray} 
The mass $M_{\zeta^{\pm}_{L}}$ and $M_{\zeta^{\pm}_{R}}$ corresponding to $\zeta^{\pm}_{mL}$ and $\zeta^{\pm}_{mR}$ fields can be obtained using the following relations
\begin{equation}
M^{2}_{\zeta^{\pm}_{L}/\zeta^{\pm}_{R}} = \frac{1}{2}\left[ M_{R} + M_{L} \mp \sqrt{\left( M_{R} - M_{L} \right)^{2} - 4M^{2}_{LR}} \right]
\end{equation}
\noindent
In Eq.\,\ref{Eq:Hc_M_Eigen} we present the physical basis of the singly charged scalars in terms of gauge basis.  
\begin{eqnarray}
\chi_m^\pm &\sim& \chi_m^\pm \nonumber \\
\zeta_{mL}^\pm & = &\cos\theta^\pm~\zeta_{L}^\pm + \sin\theta^\pm~\zeta_{R}^\pm \nonumber \\
\zeta_{mR}^\pm & = & - \sin\theta^\pm~\zeta_{L}^\pm + \cos\theta^\pm~\zeta_{R}^\pm
\label{Eq:Hc_M_Eigen}
\end{eqnarray}

\subsection{Fermion Sector}\label{SubSec:Fermion}	 
\noindent

The matter sector of the model contains both the left and right handed doublet fermions which are transforming under the $SU(2)_L$ and $SU(2)_R$ gauge group respectively. In addition to that there are $SU(2)$ singlet fields correspond to each of the charged fermions. 
\begin{eqnarray}
{Q^{i}_{L,R}} = 
\begin{bmatrix}
	u^{i}_{{L,R}}\\d^{i}_{{L,R}}
\end{bmatrix},
~~{L^{i}_{L,R}}  = 
\begin{bmatrix}
	\nu^{i}_{{L,R}}\\e^{i}_{{L,R}}
\end{bmatrix},
~~U^{i}_{L,R},
~~D^{i}_{L,R},
~~E^{i}_{L,R}
\label{Eq:fermionrep}
\end{eqnarray}
\noindent
In Eq.\,\ref{Eq:fermionrep} $Q^{i}_{L}$($L^{i}_{L}$) represent $SU(2)_L$ doublets and $SU(2)_R$ singlets whereas $Q_{R}$($L_{R}$) are $SU(2)_R$ doublets and $SU(2)_L$ singlets, respectively. $U^{i}_{L,R},D^{i}_{L,R}$ and $E^{i}_{L,R}$ are the heavy fermionic fields that transform as singlets \emph{w.r.t} $SU(2)_L \times SU(2)_R$ gauge group. The index i$ \in $[1,3] signifies different generations of these fermions. The gauge charge assignment correspond to each of these fermion fields is illustrated in Table.\,\ref{Tab:Fermion}. 

 \FloatBarrier
	 \begin{table}[H]
	 	\centering
	 	\resizebox{0.5\textwidth}{!}{
		\begin{tabular}{|c|c|}
			\hline
			Multiplets & $SU(3)_C\times SU(2)_R\times SU(2)_L\times U(1)_{B-L}$ \\
			\hline
			\hline
			Quarks&
			\pbox{10cm}{
				\vspace{2pt}
				${Q^i_L} (3,1,2,\frac{1}{3})$ = 
				$\begin{bmatrix}
					u^{i}_{L}\\d^{i}_{L}
				\end{bmatrix}$\\
				${Q^{i}_R} (3,2,1,\frac{1}{3})$ = 
				$\begin{bmatrix}
				u^{i}_{R}\\d^{i}_{R}
				\end{bmatrix}$\\
				$U^{i}_{L,R}(3,1,1,\frac{4}{3})$\\
				$D^{i}_{L,R}(3,1,1,\frac{-2}{3})$\\
				\vspace{2pt}}\\
			\hline
			\hline
				Leptons&
				\pbox{10cm}{
					\vspace{2pt}
					${L_L}_i (1,1,2,-1)$ = 
					$\begin{bmatrix}
					\nu_{L_i}\\e_{L_i}
					\end{bmatrix}$\\
					${L_R}_i (1,2,1,-1)$ = 
					$\begin{bmatrix}
					\nu_{R_i}\\e_{R_i}
					\end{bmatrix}$\\
					$E_{{L,R}_i}(1,1,1,-2)$\\
					\vspace{2pt}}\\
\hline
\end{tabular}
	    }
    \caption{ 
Fermions content of the model
    }\label{Tab:Fermion}
	\end{table}
\noindent
The presence of both the left and right fermion doublets is an immediate consequence of the Left-Right symmetric nature of this model. In absence of neutral singlet fermion field, the model prohibits neutrinos to get mass
in the same way as the charged fermions do at tree level. In Eq.\,\ref{Eq:LFermion} we write down the fermion Lagrangian following the gauge symmetry as. 

\begin{eqnarray}
	\mathcal{L}_{yuk} &=& ~ \mathcal{Y}^{ij}_{uL} \overline{Q}_{Li}\tilde{\Phi}_L U_{Rj} +\mathcal{Y}^{ij}_{dL} \overline{Q}_{Li}\Phi_L D_{Rj}+\mathcal{Y}^{ij}_{eL} \overline{L}_{Li}\Phi_L E_{Rj} + \mathcal{Y}^{ij}_{z} \overline{L^c_{Li}}\zeta_{L}E_{Lj} + \mathcal{Y}^{ij}_{c}\overline{L^c_{Li}} L_{Lj}\chi^{+}  \nonumber \\
	& +& \mathcal{Y}^{ij}_{q}\overline{U_{L}}D_{R}\chi^{+} + (L\rightarrow R)  + \overline{U_{Li}}\mathcal{M}^{ij}_{U} U_{Rj} +\overline{D_{Li}}\mathcal{M}^{ij}_{D} D_{Ri}+\overline{E_{Li}}\mathcal{M}^{ij}_{E} E_{Rj} +h.c. 
	\label{Eq:LFermion}   
\end{eqnarray}
\noindent
Here $i, j$ are the generation indices and $\tilde{\Phi}_{L/R} = i\tau_{2}\Phi^{*}_{L/R}$ where $\tau_{2}$ stands for second Pauli matrix. Apart from the $\Phi_{L/R}$ fields, the other scalar representations, \emph{viz} $\zeta_{L,R}$ and $\chi^\pm$ do not contribute to different charged fermion masses. However, they play important role in neutrino mass through loop mediated processes. We will  discuss this elaborately in Sec.\,\ref{Sec:MassNu}. Before writing down the mass matrices correspond to quarks and leptons we like to comment on the different parameters that are involved in Eq.\,\ref{Eq:LFermion}.  

\begin{itemize}
\item To obtain correct SM fermion masses we consider that the Yukawa couplings $\mathcal{Y}^{ij}_{uL}$/$\mathcal{Y}^{ij}_{dL}$/$\mathcal{Y}^{ij}_{eL}$ resemble with the SM counterpart. We further assume that all these Yukawa matrices are real and do not give rise to CP-violating interactions. The CP-violation is not the focus of our current study. Due to this consideration the couplings between Higgs and SM-like fermions receive minimal modification.
     
\item To make our calculation simple we consider that the bare mass terms correspond to the heavy vector like fermions, $\mathcal{M}^{ij}_{U/D/E}$, are diagonal. As an outcome these matrices can not serve as a possible source for FCNC processes. 

\item To simplify our calculation even further, we fix the explicit form of  $\mathcal{Y}^{ij}_{z}$ and $\mathcal{Y}^{ij}_{c}$ matrices. From space-time symmetry one can realise that the $\mathcal{Y}^{ij}_{c}$ is anti-symmetric matrix. However to minimize the number of independent parameters we considered the $\mathcal{Y}^{ij}_{z}$ matrix to be diagonal which is in reality can be an arbitrary $3\times3$ complex matrix. Keeping that in mind we define the matrix $\mathcal{Y}^{ij}_{z}$ and $\mathcal{Y}^{ij}_{c}$ in following fashion.   
\begin{eqnarray}
\mathcal{Y}^{ij}_{z} \subset \{\text{Symmetric Matrices}\} \rightarrow \begin{bmatrix} \mathcal{Y}_{11} & 0 & 0 \\
	0 & \mathcal{Y}_{22} & 0 \\
	0 & 0 & \mathcal{Y}_{33} \end{bmatrix}
\label{Eq:yz}
\end{eqnarray} 
\begin{eqnarray} \mathcal{Y}^{ij}_{c} \subset \{\text{Anti-Symmetric Matrices}\} \rightarrow  \begin{bmatrix}0 & \mathcal{Y}^{c}_{12} & \mathcal{Y}^{c}_{13} \\
-\mathcal{Y}^{c}_{12} & 0 & \mathcal{Y}^{c}_{23} \\
-\mathcal{Y}^{c}_{13} & -\mathcal{Y}^{c}_{23} & 0 \end{bmatrix}
\label{Eq:yc}
\end{eqnarray}
\end{itemize}
\begin{figure}[h]
	\centering
	\includegraphics[height=2.6cm,width=0.3\textwidth]{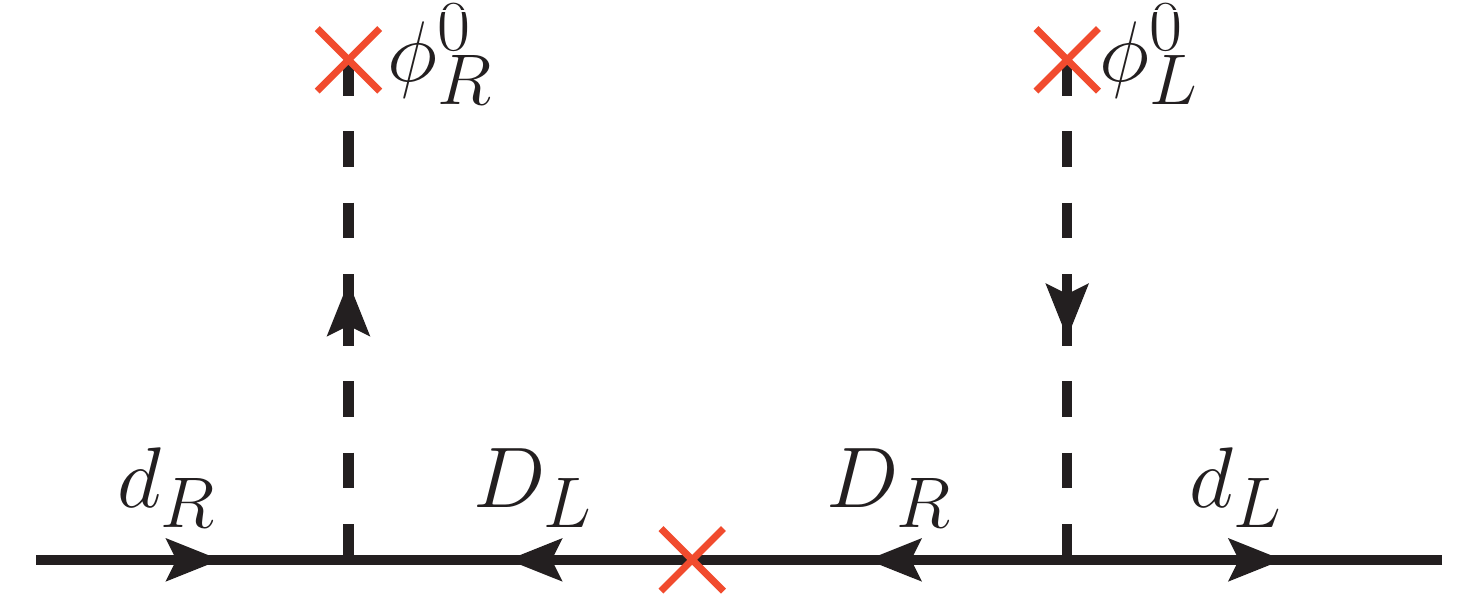}
	\includegraphics[height=2.6cm,width=0.3\textwidth]{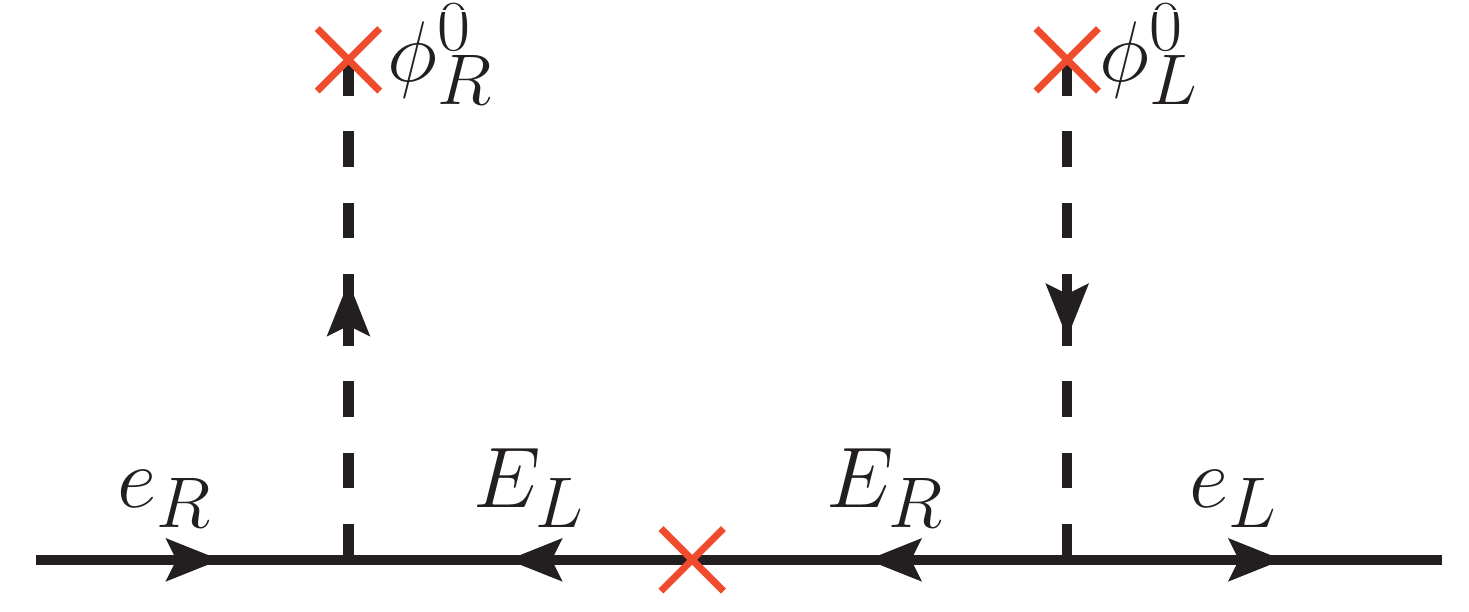}
	\includegraphics[height=2.6cm,width=0.3\textwidth]{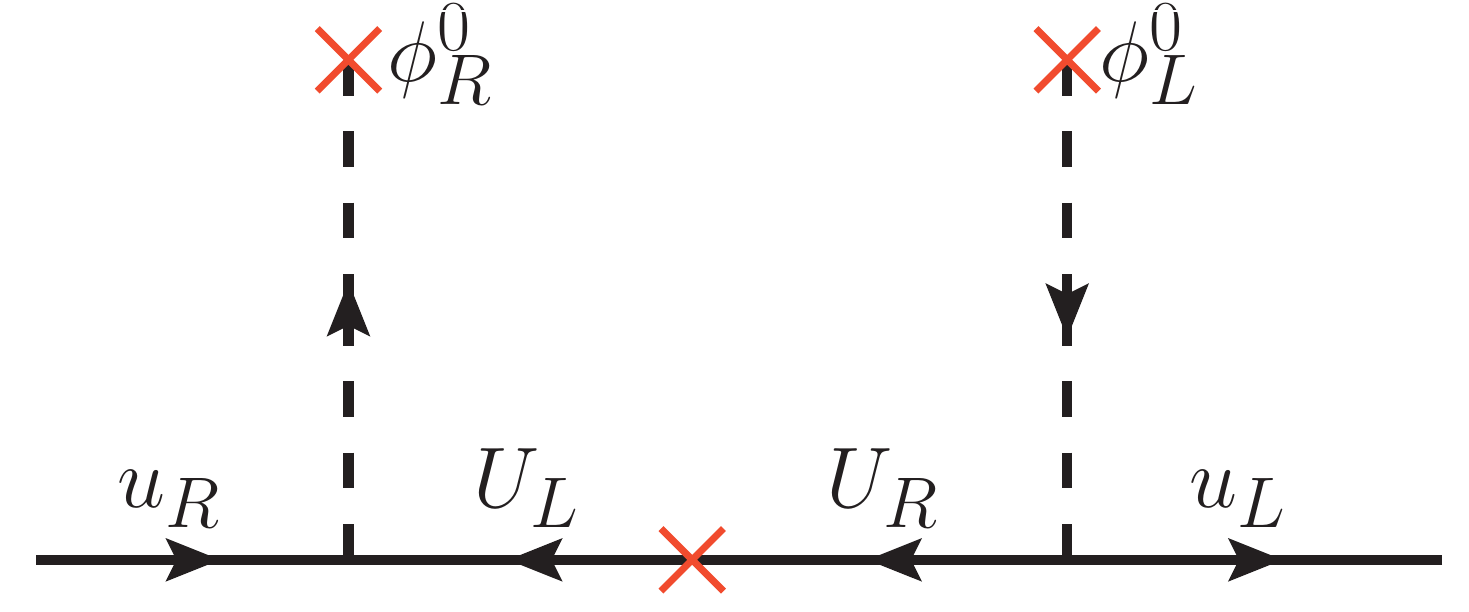}
	\caption{Generation of fermion masses through universal seesaw mechanism}
	\label{Fig:Chragedfermion_mass}
\end{figure}

\noindent
Invoking symmetry breaking in Eq.\,\ref{Eq:LFermion} one can write down the mass matrices correspond to the charged fermions as discussed in  \cite{Deppisch:2016scs,Davidson:1987mh,Dev:2015vjd,Babu:1988mw}
\begin{equation}
M_{u} = \begin{bmatrix}
0 & \frac{\mathcal{Y}_{uL}v_{L}}{\sqrt{2}} \\
\frac{\mathcal{Y}^{\dagger}_{uR}v_{R}}{\sqrt{2}} & \mathcal{M}_{U}
\end{bmatrix},~~~ M_{d} = \begin{bmatrix}
0 & \frac{\mathcal{Y}_{dL}v_{L}}{\sqrt{2}} \\
\frac{\mathcal{Y}^{\dagger}_{dR}v_{R}}{\sqrt{2}} & \mathcal{M}_{D}
\end{bmatrix},~~~ M_{e} = \begin{bmatrix}
0 & \frac{\mathcal{Y}_{eL}v_{L}}{\sqrt{2}} \\
\frac{\mathcal{Y}^{\dagger}_{eR}v_{R}}{\sqrt{2}} & \mathcal{M}_{E}
\end{bmatrix}
\label{Eq:Mx_Fermion} .
\end{equation}

\begin{figure}[h]
	\centering
	\includegraphics[height=8.5cm,width=0.65\textwidth]{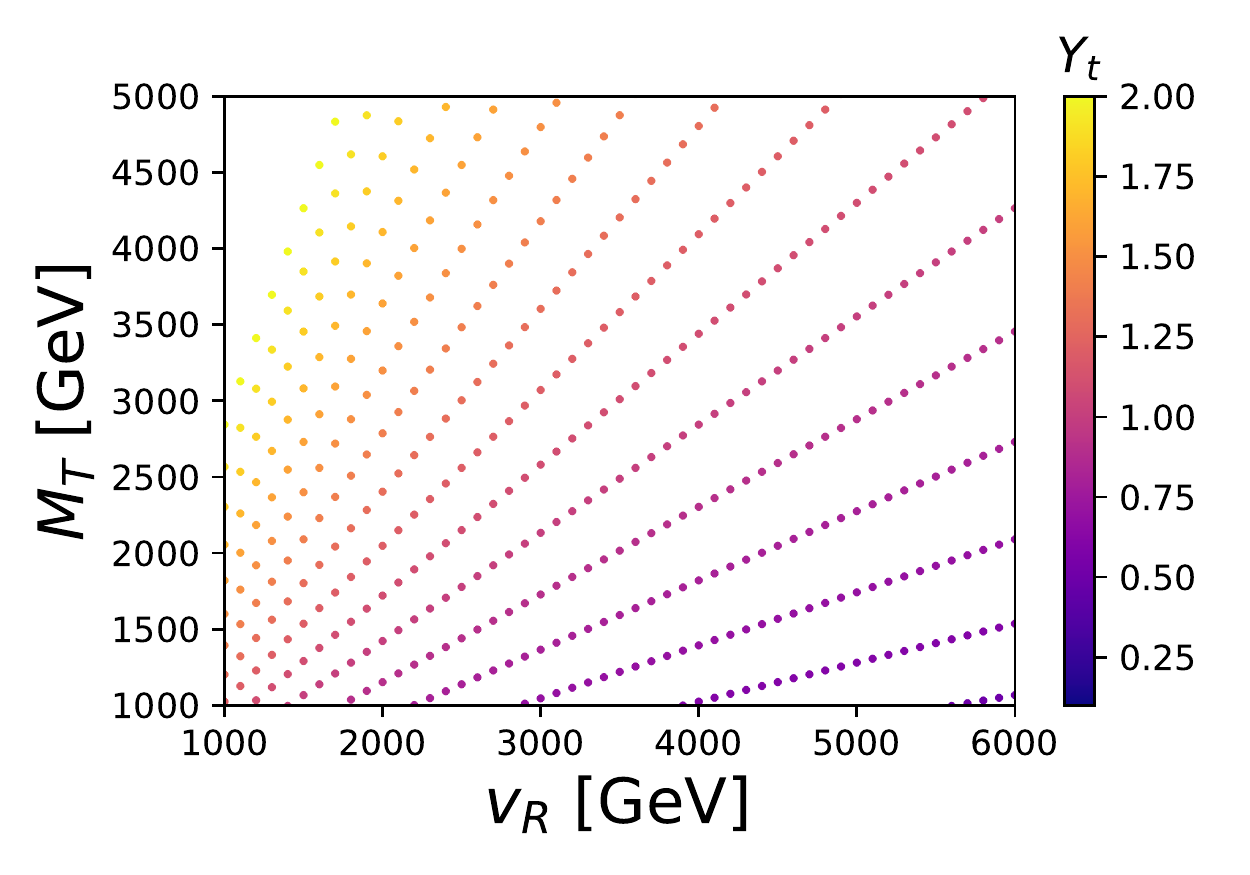}
	\caption{Variation of $heavy-top$  mass ($\mathcal{M}_T$) as a function of ($v_R$) for different values of top Yukawa coupling $\mathcal{Y}_t$. }
	\label{Fig:MT_yt}
\end{figure}

\noindent
The generation of fermion masses is
diagrammatically illustrated in Fig.\,\ref{Fig:Chragedfermion_mass}. To obtain the mass eigenstates, corresponding to mass matrices that are mentioned in Eq.\,\ref{Eq:Mx_Fermion} one needs to diagonalise these using appropriate bi-unitary transformations.  
\begin{eqnarray}
	M^{D}_{x} = U^{L}_{x}M_{x}U^{R\dagger}_{x}
\label{Eq:Unitaryfermion}
\end{eqnarray}
In Eq.\,\ref{Eq:Unitaryfermion} we present the diagonalisation of the mass matrices while invoking unitary rotations. Here the subscript $x \in [u,d,e]$ for up-quark, down-quark and charged lepton sector respectively. These unitary matrices, $U^{L/R}_{x}$, can be parameterised as, 
\begin{equation}
	U^{L/R}_{x} = \begin{bmatrix}
		1 - \frac{\rho^{L/R}_{x}\rho^{\dagger L/R}_{x}}{2} & - \rho^{L/R}_{x} \\
		\rho^{\dagger L/R}_{x} & 1 - \frac{\rho^{\dagger L/R}_{x}\rho^{L/R}_{x}}{2}
	\end{bmatrix}
	\label{Eq:U_Fermion}
\end{equation}
where $ \rho^{L}_{x} = \frac{v_{L}\mathcal{Y}_{x}}{\mathcal{M}_{X}},~~ \rho^{R}_{x} = \frac{v_{R}\mathcal{Y}_{x}}{\mathcal{M}_{X}}$ and $\mathcal{M}_{X}$ with $X \in [U,D,E]$ are 3$\times$3 diagonal matrices. If we assume the bare mass terms of the vector-like fermions to be sufficiently large, then in the limit $v_{L}\mathcal{Y}_{x}\ll v_{R}\mathcal{Y}_{x}\ll \mathcal{M}_{X}$, the elements of $\rho_{L,R}$ will be significantly smaller than unity. Using the matrices $U^{L/R}_{x}$ as described in Eq.\,\ref{Eq:U_Fermion}, the mass matrices can be diagonalised and the standard model charged fermion masses would take the following form,

\begin{eqnarray}
	m_u \simeq \frac{v_{L}v_{R}\mathcal{Y}^2_{u}}{\mathcal{M}_{U}},~~m_d \simeq \frac{v_{L}v_{R}\mathcal{Y}^2_{d}}{\mathcal{M}_{D}},~~m_e \simeq \frac{v_{L}v_{R}\mathcal{Y}^2_{e}}{\mathcal{M}_{E}}
	\label{Eq:seesaw_reln}
\end{eqnarray}
\noindent
Such seesaw realisation of fermion masses are termed as \textit{Universal seesaw mechanism} \cite{Davidson:1987mh,Mohapatra:1987nx,Ma:1989tz,Ma:2017kgb,Brahmachari:2003wv}. The aforementioned seesaw relation has an interesting consequence on the $heavy-top$ mass, $\mathcal{M}_T$. From Eq.\,\ref{Eq:seesaw_reln} one can notice, if $\mathcal{M}_T >> v_R$, then $\mathcal{Y}_t$ should be much larger than one to satisfy correct top quark mass. For this large value of $\mathcal{Y}_t$ the underlying model would violate perturbativity and consequently leading to instability of the electroweak vacuum. In Fig.\,\ref{Fig:MT_yt}, we have shown the variation of the $heavy-top$ mass as a function of $v_R$ to show the variation of the top Yukawa coupling ($\mathcal{Y}_t$). This will help us to identify the allowed region parameter space of our model which is  $\mathcal{M}_T \leq v_R$.

\section{Neutrino Mass Generation}\label{Sec:MassNu}
\begin{figure}[H]
	\centering
	\includegraphics[height=4.5cm,width=0.4\textwidth]{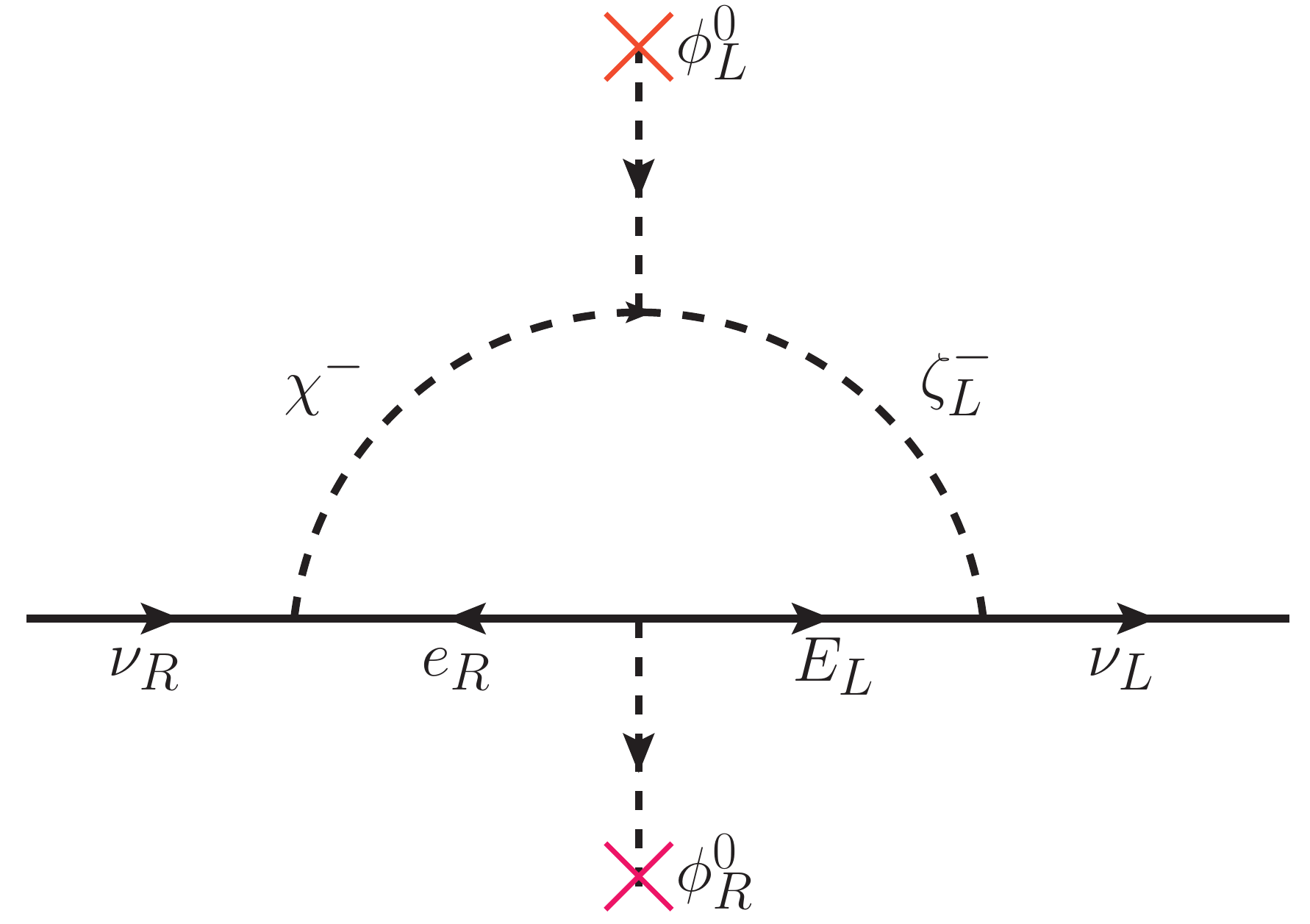}
	\includegraphics[height=4.5cm,width=0.4\textwidth]{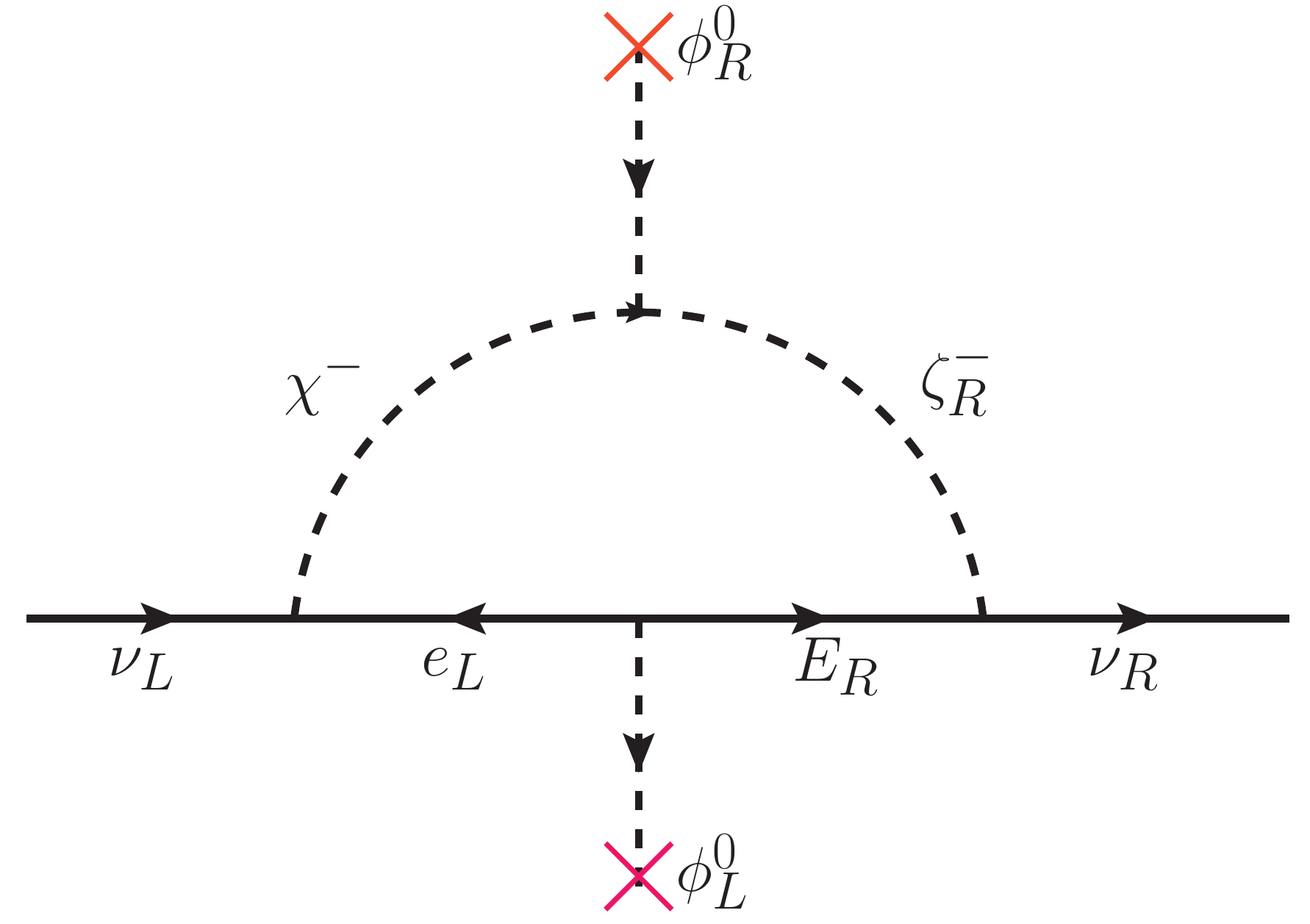}
	\caption{Diagrams responsible for One-loop neutrino mass generation in gauge basis.}
	\label{Fig:NeutrinoG}
\end{figure}

\begin{figure}[H]
	\includegraphics[height=3cm,width=0.4\textwidth]{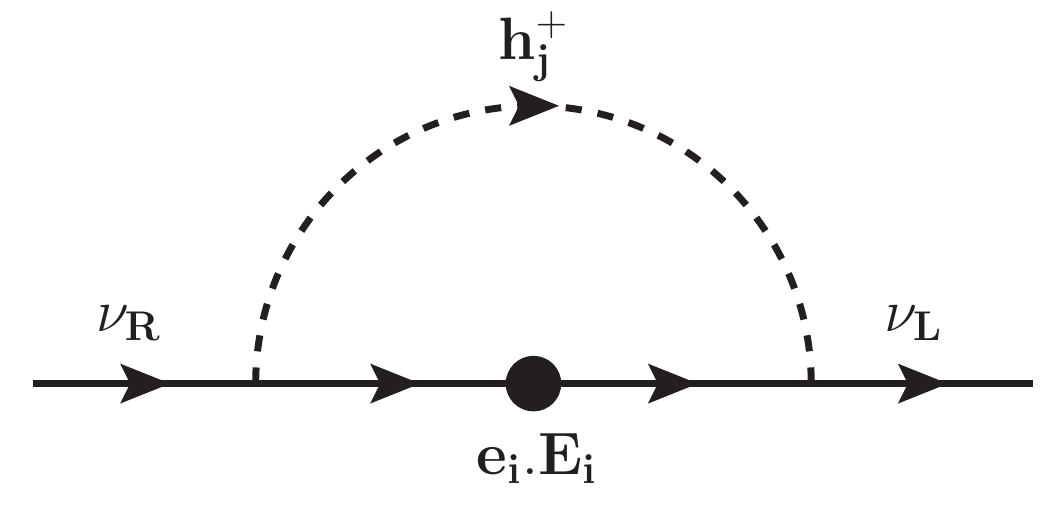}
	\hspace{0.5cm}
	\includegraphics[height=3cm,width=0.4\textwidth]{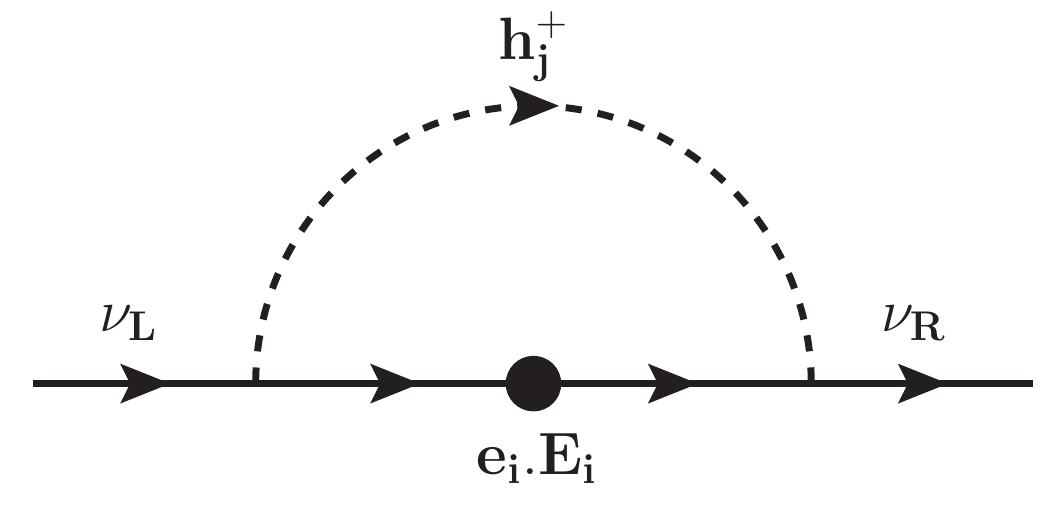}
	\caption{Diagrams responsible for One-loop neutrino mass generation in mass basis.}
	\label{Fig:NeutrinoM}
\end{figure}

In the last section we have laid out the essential details of the model. Now we will turn our attention to neutrino mass generation. Due to the absence of gauge singlet field components, the neutrinos in this model do not achieve the necessary masses via tree level Yukawa like terms. However, the additional lepton and scalar fields of the model  that are charged under the gauge group $SU(2)_{L}\times SU(2)_{R}\times U(1)_{B - L}$ interact via loop mediated process with neutrinos  and induce small Dirac type neutrino mass. The model conserve lepton number which is the consequence of the Dirac neutrinos present in our model.  To elaborate this point, let us write down the essential part of the fermion sector Lagrangian, as discussed in Eq.\,\ref{Eq:LFermion}.   
\begin{equation}
\mathcal{L}_{fermion} \subset \mathcal{Y}_{zij}\overline{L^{c}_{Li}}\zeta_{L}E_{Lj} + \mathcal{Y}_{cij}\overline{L^{c}_{Li}}L_{Lj}\chi^{+} + \left(L \rightarrow R\right) + h.c.
\label{Eq:Lneutrino}
\end{equation}
\noindent
Expanding the above Eq.\,\ref{Eq:Lneutrino}, one can figure out all possible Feynman diagrams that would generate the one loop neutrino mass. In Fig.\,\ref{Fig:NeutrinoG} we present the corresponding diagrams in the gauge basis. The right plot of  Fig.\,\ref{Fig:NeutrinoG} can be obtained from the left plot by simply interchanging the left and right handed components respectively. In the mass basis of the fermions the left plot give two contributions, \emph{i.e.} the contribution of $e$ and $E$, to neutrino mass: $(m_\nu(e) + m_\nu(E))\bar{\nu}_L\nu_R$ + h.c., similarly the right plot also has two contribution: $(\tilde{m}^\dagger_\nu(e) + \tilde{m}^\dagger_\nu(E))\bar{\nu}_R\nu_L$ + h.c. Here the flavor indices are suppressed. Hence there are four contributions, two from each diagrams, which would generate neutrino mass at one loop in our model. This can be clearly seen from Fig.\,\ref{Fig:NeutrinoM}

In Fig.\,\ref{Fig:NeutrinoM} we present the loop diagrams contributing to the neutrino mass generation considering the mass basis of different scalar and charged fermions. Both the light and heavy charged leptons along with the heavy singly charged scalars are responsible for radiative one loop Dirac mass of neutrinos. Considering all the contributions the 3$\times$3 non-diagonal neutrino mass matrix in the flavor basis, \emph{viz} $\nu_e$, $\nu_\mu$, $\nu_\tau$ will be,
 
\begin{equation}
	m_{\nu}^{mn} = m_{\nu}\left(e_{i}\right)^{mn} + m_{\nu}\left(E_{i}\right)^{mn} + \tilde{m}_{\nu}\left(e_{i}\right)^{mn} + \tilde{m}_{\nu}\left(E_{i}\right)^{mn} 
	\label{Eq:Nu_Mass}  
\end{equation}

where m,n $ \in $ [$e,\mu,\tau$] for three flavors of neutrinos. Each of the individual contributions to the neutrino mass is obtained as, 

\begin{eqnarray}
	m_{\nu}\left(e_{i}\right)^{mn}  &=& \mathcal{Y}^{mi}_{c}\left[U^{Lf}_{11}\frac{m_{e_{i}}}{16\pi^{2}}\sum_{j}U^{h^{\pm}}_{1j}U^{h^{\pm}*}_{3j}\frac{m^{2}_{h^{\pm}_{j}}}{m^{2}_{h^{\pm}_{j}} - m^{2}_{e_{i}}}\ln\left(\frac{m^{2}_{h^{\pm}_{j}}}{m^{2}_{e_{i}}}\right)U^{Rf\dagger}_{21}\right]\mathcal{Y}^{ni}_{z} \nonumber \\
	m_{\nu}\left(E_{i}\right)^{mn} & =& \mathcal{Y}^{mi}_{c}\left[U^{Lf}_{12}\frac{m_{E_{i}}}{16\pi^{2}}\sum_{j}U^{h^{\pm}}_{1j}U^{h^{\pm}*}_{3j}\frac{m^{2}_{h^{\pm}_{j}}}{m^{2}_{h^{\pm}_{j}} - m^{2}_{E_{i}}}\ln\left(\frac{m^{2}_{h^{\pm}_{j}}}{m^{2}_{E_{i}}}\right)U^{Rf\dagger}_{22}\right]\mathcal{Y}^{ni}_{z} \nonumber \\
	\tilde{m}_{\nu}^\dagger\left(e_{i}\right)^{mn} & =& \mathcal{Y}^{mi}_{c}\left[U^{Rf}_{11}\frac{m_{e_{i}}}{16\pi^{2}}\sum_{j}U^{h^{\pm}}_{1j}U^{h^{\pm}*}_{2j}\frac{m^{2}_{h^{\pm}_{j}}}{m^{2}_{h^{\pm}_{j}} - m^{2}_{e_{i}}}\ln\left(\frac{m^{2}_{h^{\pm}_{j}}}{m^{2}_{e_{i}}}\right)U^{Lf\dagger}_{21}\right]\mathcal{Y}^{ni}_{z} \nonumber \\
	\tilde{m}_{\nu}^\dagger\left(E_{i}\right)^{mn} & =& \mathcal{Y}^{mi}_{c}\left[U^{Rf}_{11}\frac{m_{E_{i}}}{16\pi^{2}}\sum_{j}U^{h^{\pm}}_{1j}U^{h^{\pm}*}_{2j}\frac{m^{2}_{h^{\pm}_{j}}}{m^{2}_{h^{\pm}_{j}} - m^{2}_{E_{i}}}\ln\left(\frac{m^{2}_{h^{\pm}_{j}}}{m^{2}_{E_{i}}}\right)U^{Lf\dagger}_{21}\right]\mathcal{Y}^{ni}_{z}
	\label{Eq:Loop_Nu}
\end{eqnarray}

where, $U^{L,R}$ are the charged lepton mixing matrix, $U^{h^\pm}$ is the singly charged scalar mixing matrix and $m_{h^{\pm}_{j}}$ $ \in $ [$M_{\chi^\pm}$,$M_{\zeta_L^\pm}$,$M_{\zeta_R^\pm}$]. For better understanding of the aforementioned contributions to the neutrino mass we have elaborately discussed one contribution, the contribution of SM like leptons, in Appendix[\ref{App:One_loop}].



%

The 3$\times$3 non-diagonal matrix in Eq.\,\ref{Eq:Nu_Mass} can be diagonalised by bi-unitary transformation to generate the light neutrino mass in mass basis, \emph{viz} $\nu_1,\nu_2,\nu_3$ as,
\begin{equation}
U^{\dagger}m_{\nu}V = m^{d}_{\nu}
\label{Eq:Nu_diag}
\end{equation}
\noindent
where $U$ is the usual PMNS matrix and $V$ is the right-handed counterpart. 

\begin{figure}[h]
\centering
	\includegraphics[height=4cm,width=0.4\textwidth]{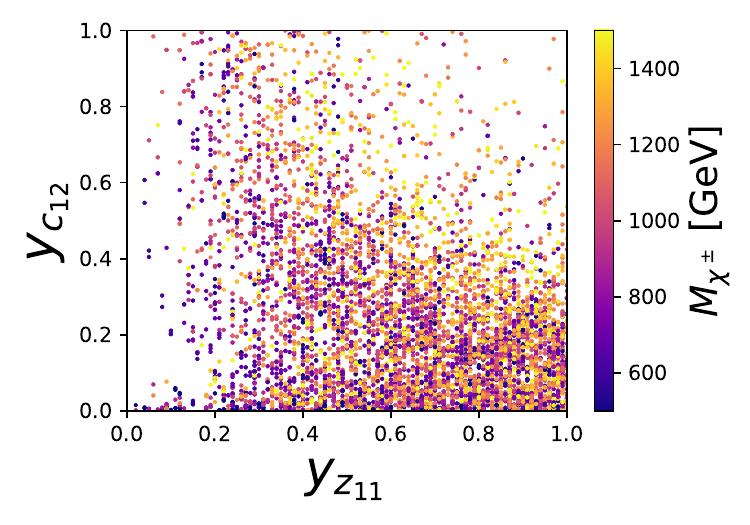}
	\includegraphics[height=4cm,width=0.4\textwidth]{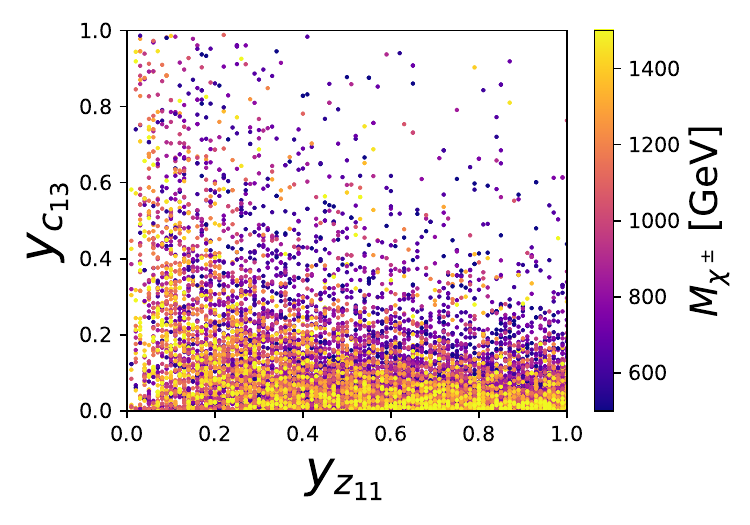}
	\includegraphics[height=4cm,width=0.4\textwidth]{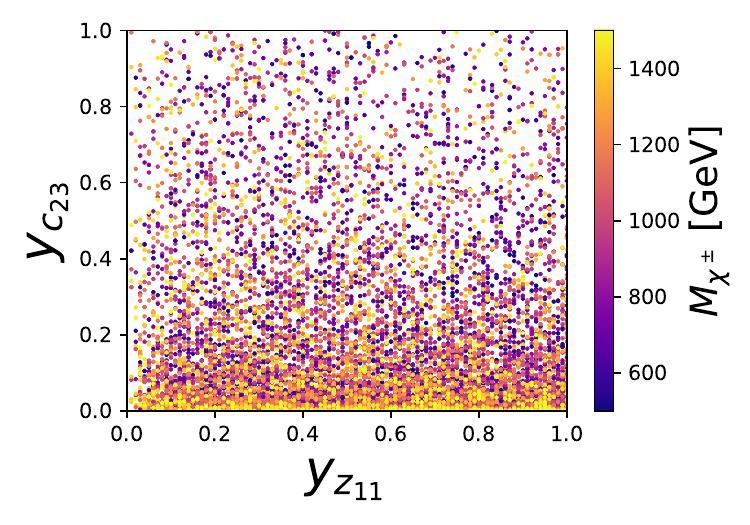}
	\includegraphics[height=4cm,width=0.4\textwidth]{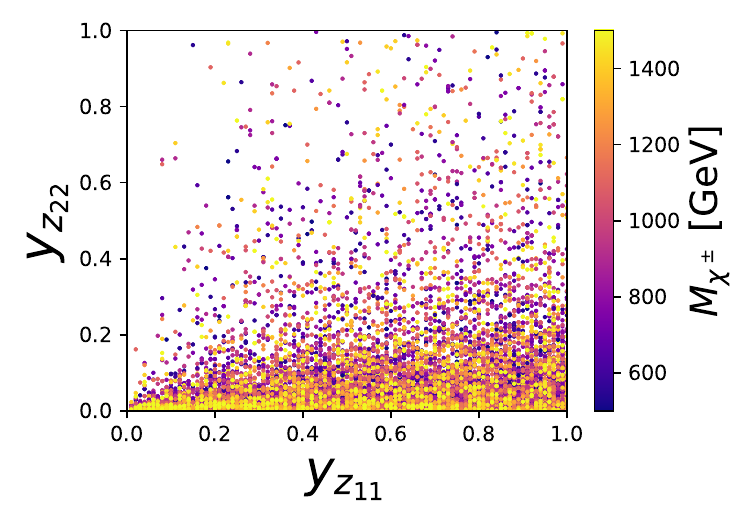}
	\includegraphics[height=4cm,width=0.4\textwidth]{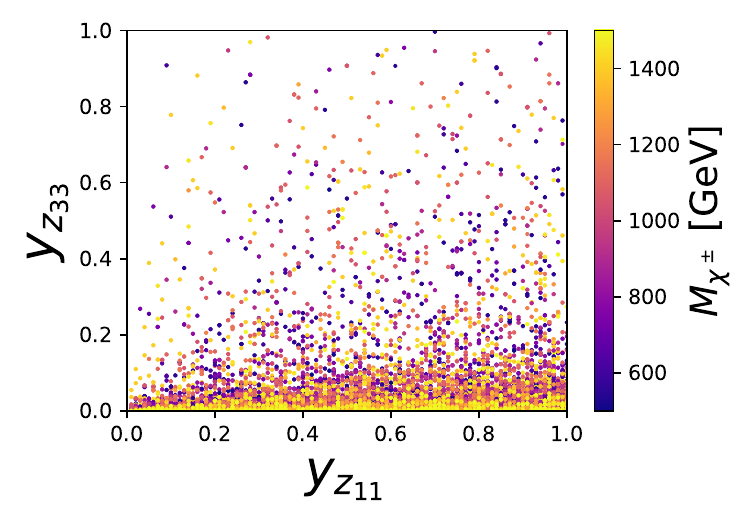}
	\caption{The figure shows the variation of Yukawa couplings, $ \mathcal{Y}_{z} $ and $ \mathcal{Y}_{c} $, with respect to $M_\chi^\pm$ taking $3\sigma$ variation in  neutrino oscillation data with an exception of $\delta_{CP} = 180^o$ in Normal hierarchy. The values of masses of other heavy particles are mention in the text. The figure  proves the existence of available parameter space in our model. Similar behavior has been obtained for Inverted hierarchy and hence not shown separately.  }
	\label{Fig:NO}
\end{figure}

In Fig.\,\ref{Fig:NO} we have shown the variation of Yukawa couplings, $ \mathcal{Y}_{z} $ and $ \mathcal{Y}_{c} $, with respect to the singly charged scalar  mass $M_\chi^\pm$ for normal hierarchy. For inverted hierarchy the behavior is same as that of normal hierarchy and hence we have not shown them separately. The figures provide evidence for the existence of parameter space satisfying $3\sigma$ neutrino oscillation data. Assuming $M_\chi^\pm$ to be the lightest one, hence considered to be an independent variable, the heavy vector like leptons and the heavy charged scalars are considered to be of the $\mathcal{O}[\text{TeV}]$: $M_E = $ 10 TeV, $M_{\zeta_L^\pm} =  2 M_{\chi\pm}$ and  $M_{\zeta_R^\pm} =  3 M_{\chi\pm}$. Among the neutrino oscillation parameters, for the CP-phase we consider $\delta_{CP} = 180^o$. From Eq.\,\ref{Eq:Loop_Nu}, It is clear that out of the two contributions: the SM leptonic contribution and the heavy leptonic contribution, the major contribution comes from the heavy fermion loop. \emph{i.e.} $m_{E}\times\frac{m^{2}_{h^{\pm}}}{m^{2}_{h^{\pm}} - m^{2}_{E}}\ln\left(\frac{m^{2}_{h^{\pm}}}{m^{2}_{E}}\right)$. Also for a given heavy fermion mass, with increase in charged scalar mass increases the corresponding loop contribution. From Fig.\,\ref{Fig:NO} it is clear that with large charged scalar mass we require small diagonal Yukawa couplings, \emph{viz} $\mathcal{Y}_z$, however the anti-symmetric Yukawas $\mathcal{Y}_c$ don't follow a particular pattern and can span over a large range of parameter space.

\section{\label{Sec:Pheno}Phenomenology}
We now turn our attention towards the phenomenological aspects of this model with the extended particle content: the additional heavy gauge bosons, scalar particles as well as fermions. With an extensive charged as well as neutral scalar spectrum, the model has a rich phenomenological implication. Significant number of searches for BSM scalars, both charged and neutral, have been carried out at the LHC by ATLAS and CMS collaboration, and the scarcity of any excess  over SM signal ensued stringent constraints on viable parameter space. Firstly we discuss some of the important constraints applicable on the model parameter space and then quantify production cross-section of some of the BSM particles in pursuit of the future collider searches.

\subsection{Constraints}
\subsubsection{Constraints from LFV}

The presence of doubly charged Higgs and its interaction with the SM fermions can give rise to possible lepton flavour violating (LFV) processes. A detailed discussion on LFV constraints can be found in \cite{BhupalDev:2018tox,Akeroyd:2009nu,Dev:2017ouk}. From the fermion sector Lagrangian as described in Eq.\,\ref{Eq:LFermion}, one can notice that no two charged SM leptons couple to the doubly charged Higgs (Eq.\,\ref{Eq:Lneutrino} for details) through the tree level interaction terms. Therefore processes such as $l_i \rightarrow l_j l_k l_l$ can not occur at tree level. However the aforementioned process and other LFV processes, such as $\mu\rightarrow e\gamma$ can occur at one-loop where light-heavy charged lepton mixing play a crucial role.  Among the limits that are obtained from non-observation of different LFV processes, we consider the following stringent constraints for
$\mu \rightarrow e e e $ and $\mu \rightarrow$  $e\gamma$. The current limits on these are BR($\mu \rightarrow e e e $) $<$ $1.0 \times 10^{-12}$~ \cite{SINDRUM:1987nra} and BR($\mu \rightarrow e \gamma$) $<$ $4.2 \times 10^{-13}$~\cite{MEG:2016leq}. In our model these constraints lead to following relations between the doubly charged Higgs mass as well as relevant Yukawa terms. 
\begin{eqnarray}
	\mu \rightarrow e e e&:& |(U_{11}^L)_{2i}\mathcal{Y}^{ij}_{z}(U_{21}^L)_{j1}||(U_{11}^L)_{1\alpha}\mathcal{Y}^{\alpha\beta}_{z}(U_{21}^L)_{\beta1}| < 2.3 \times 10^{-7} \Big(\frac{m_H^{++}}{100 ~\text{GeV}}\Big)^2 \nonumber \\ 
		\mu \rightarrow e \gamma&:& |(U_{11}^L)_{2i}\mathcal{Y}^{ij}_{z}(U_{21}^L)_{jk}||(U_{11}^L)_{k\alpha}\mathcal{Y}^{\alpha\beta}_{z}(U_{21}^L)_{\beta1}| < 2.7 \times 10^{-6} \Big(\frac{m_H^{++}}{100 \,\text{GeV}}\Big)^2
\end{eqnarray} 
 Here the repeated indices are summed over. In the above, $U^{L}_{ij}$ and $\mathcal{Y}_{z}$ are the left-handed unitary rotation and the Yukawa matrix respectively. 
  In our current study, we have chosen the Yukawa matrix $\mathcal{Y}_{z}$ to be diagonal (see Eq.\,\ref{Eq:yz}). We have also considered $\mathcal{Y}_{eL}$ to be diagonal in the flavor basis, which results in a diagonal $U_{11}^L$ and $U_{21}^L$ matrix. As a consequence the LFV constraints won't affect the relevant parameter space of this model. However, if one consider a non-diagonal symmetric $\mathcal{Y}_{z}$ matrix, then the LFV constraints can still be evaded if one consider the value of different mixing parameters to be less than $\mathcal{O}(10^{-3})$. 
  \begin{figure}[H]
  	\centering
  	\includegraphics[height=8cm,width=9.5cm]{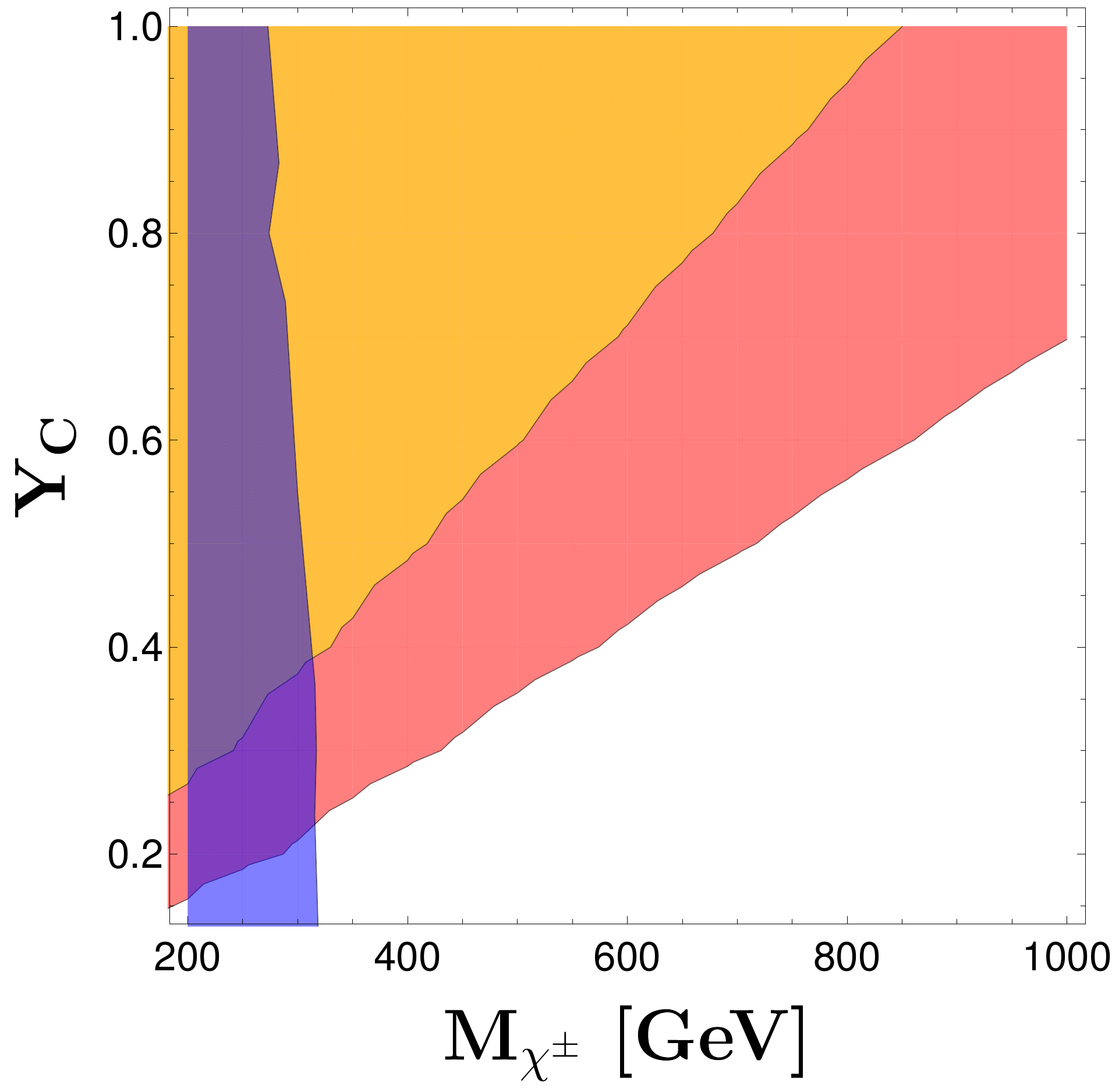}	
  	\caption{Direct as well as indirect constraints in $M_\chi^\pm$ versus $\mathcal{Y}_{c}$ plane. The blue colored region is excluded from ATLAS di-lepton+MET search \cite{ATLAS:2019lff}, and the yellow region is constrained by mono-photon search from LEP \cite{L3:2003yon}. The pink region is the excluded region from BBN \cite{Planck:2015fie}. The figure shows that the BBN constraint is more strong as compared to the ATLAS di-lepton+MET search and LEP mono-photon search.  }
  	\label{Fig:collider_bound}
  \end{figure}
\subsubsection{Collider constraints}
The doubly charged Higgs in the model can copiously be produced at the LHC. However their decay to SM particles are mixing suppressed. Similar conclusion holds for singly charged Higgs from the same multiplet $\zeta_L$ and $\zeta_{R}$. Hence the production-times-branching of the charged Higgs, mainly the scalars from the extra doublets, are not significant for LHC. However the singlet scalar $\chi^\pm$, has direct interaction with the SM leptons leading to unsuppressed branching ratios, hence the collider constraints are directly applicable on this scalar. We have reinterpreted the ATLAS dilepton+missing energy search~\cite{ATLAS:2019lff} using ChakeMate and obtained the unfavoured region marked in blue colour in Fig.\,\ref{Fig:collider_bound}. 
 
\subsubsection{Mono-photon constraints}
\begin{figure}[h]
	\centering
	\includegraphics[height=4cm,width=6cm]{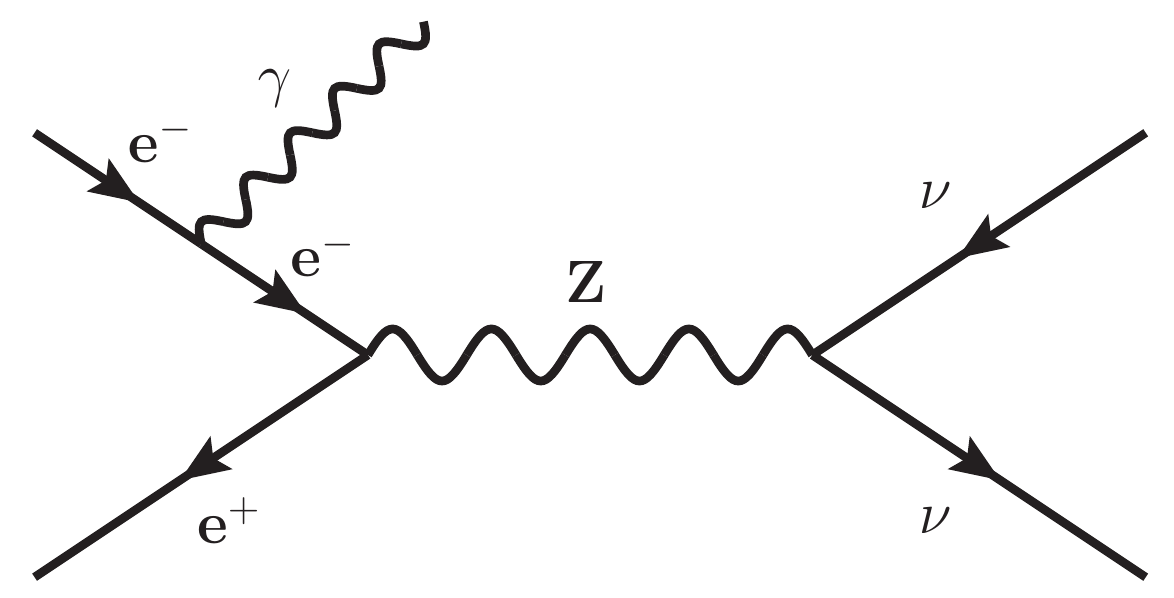}	
	\hspace{1.5cm}
	\includegraphics[height=6cm,width=4cm]{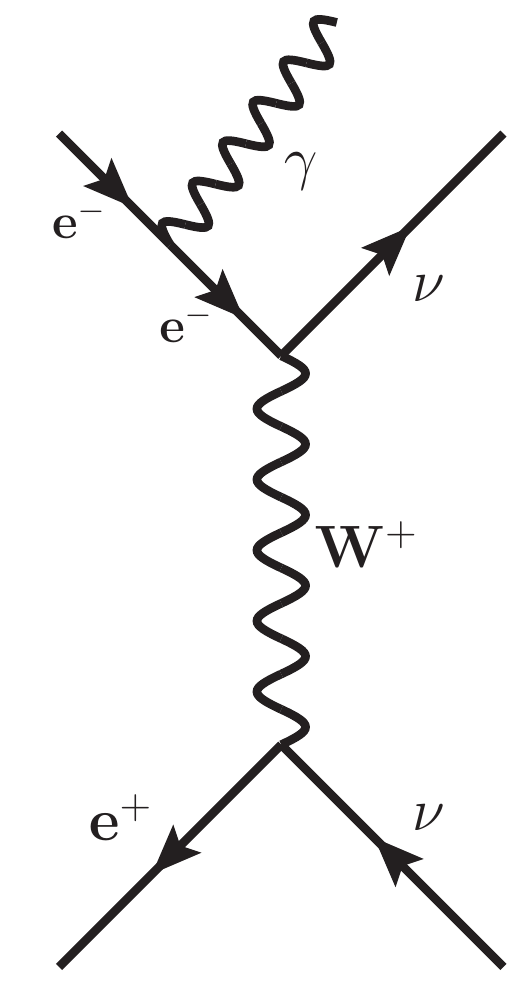}	
	\caption{Feynman diagrams of the s-channel Z boson exchange and t-channel W boson exchange for $e^{+} e^{-} \to \nu \bar{\nu} \gamma$ process in SM.}
	\label{Fig:monophotnSM}
\end{figure}

\begin{figure}[h]
	\centering
	\includegraphics[height=6cm,width=4cm]{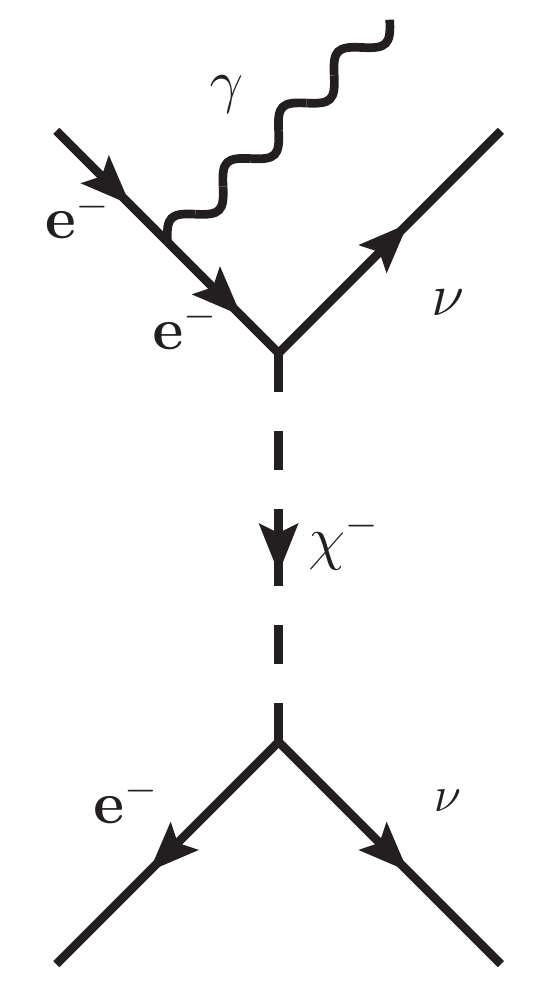}	
	\caption{Feynman diagram of the t-channel $\chi^\pm$ boson exchange for $e^{+} e^{-} \to \nu \bar{\nu} \gamma$ process.}
	\label{Fig:monophotonBSM}
\end{figure}

The singly charged Higgs $\chi^{\pm}$ primarily couples to SM leptons in this model. As a result this charged Higgs can directly contribute to $e^{+} e^{-} \to \nu \bar{\nu} \gamma$ cross section. In case of SM, the s-channel $Z$ boson exchange and t-channel $W$ boson exchange channel participate in this process. In Fig.\,\ref{Fig:monophotnSM}, we present the Feynman diagrams correspond to each of these processes. 
In presence of new physics (NP) interactions additional diagrams would contribute to this process which can potentially deviate the currently measured cross section. As a consequence, one can adopt the $\gamma + \slashed{{E}}_{T}$ cross section to impose suitable bound on the mass of the charged Higgs as well as the relevant couplings. The diagram correspond to t-channel $\chi^{\pm}$ exchange is illustrated in Fig.\,\ref{Fig:monophotonBSM}. 
The other two singly charged Higgs $\zeta^{\pm}_{L/R}$ and heavy $W_{R}$ boson also contribute to this process. As we have set their masses to a large values the effects coming from these exotic channels can be ignored owing to a large propagator suppression. To calculate the cross section we will first calculate the tree level $e^+e^- \to \nu \bar{\nu}$ cross section ($\sigma^0_{ee\rightarrow\nu\bar{\nu}}$) and incorporate the photon emission effect with an appropriate radiator function as described in \cite{Nicrosini:1988hw}. The relation between the cross section $\sigma^0_{ee\rightarrow\nu\bar{\nu}}$ and the full cross section is described in Eq.\,\ref{Eq:eenunuA}.
\begin{equation}
	\begin{split}
		\sigma_{e e \to \nu \bar{\nu} \gamma}\left(s\right) & = \int dx\int dc_{\gamma} H\left(x, s_{\gamma}; s\right)  \sigma^{0}_{e e \to \nu \bar{\nu}}\left(\hat{s}\right), \\
		H\left(x, s_{\gamma}; s\right) & = \frac{\alpha}{\pi}\frac{1}{s^{2}_{\gamma}}\frac{1 + (1 - x)^{2}}{x}
	\end{split}        
	\label{Eq:eenunuA}     
\end{equation}
\noindent
Here $s$ is the center-of-mass energy, the cross-section $\sigma^0_{ee\rightarrow\nu\bar{\nu}}$ is evaluated at the energy scale $\hat{s} = (1 - x)s$. The function $H$ determine the probability of radiation where the photon is emitting with an energy $x = \frac{2E_{\gamma}}{\sqrt{s}}$ at the angle $\theta_{\gamma}$, which is the angle between the emitted photon and the beam axis. The $s_{\gamma}$ and $c_{\gamma}$ stands for sine and cosine of the $\theta_{\gamma}$ respectively. In Eq.\,\ref{sigmamonophoton_NP}, we present the analytic expression for the $\sigma^{0}_{NP}$ that arises due to the $\chi$ exchange. The first line corresponds to the square of $\chi$ mediated amplitude whereas the second and third term signifies the interference with the $W$ and $Z$ exchange diagrams respectively.     
\begin{eqnarray}
	\sigma^{0}_{NP}\left(s\right) &=& \frac{3 \mathcal{Y}_{c}^4}{16 \pi s^2}\Bigg( \frac{(M_\chi^2+s)^2}{M_\chi^2}-2 (M_\chi^2+s) \log\Big[\frac{M_\chi^2+s}{M_\chi^2}\Big]-M_\chi^2\Bigg)\nonumber \\
	&+& \frac{G_F\mathcal{Y}_{c}^2M_W^2}{2\sqrt{2}\pi s^2(M_\chi^2-M_W^2)}\Bigg(2(M_\chi^2-M_W^2) -  (M_\chi^2 + s)^2\log\Big[\frac{M_\chi^2 + s}{M_\chi^2}\Big] + (M_W^2 + s)^2\log\Big[\frac{M_W^2 + s}{M_W^2}\Big]\Bigg)\nonumber \\
	&+& \frac{3G_F\mathcal{Y}_{c}^2M_W^2(M_Z^2-s)}{4\sqrt{2}\pi s^2((M_Z^2-s)^2+M_Z^2\Gamma_Z^2)}\Big(\frac{1-2s_w^2}{1-s_w^2}\Big)\Bigg(\frac{4sM_\chi^2+6s^2}{4}-(M_\chi^2+s)^2\log\Big[\frac{M_\chi^2+s}{M_\chi^2}\Big]\Bigg)
	\label{sigmamonophoton_NP}
\end{eqnarray}
Here $\mathcal{Y}_{c}$ and $M_{\chi}$ signifies the relevant coupling and the mass respectively. The logarithm dependency arises because of the $\chi^{\pm}$ mediation via t-channel mode.  The SM counterpart $\sigma^{0}_{SM}$ for the tree level $e^{+} e^{-} \to \nu \bar{\nu}$ process is mentioned in \cite{Berezhiani:2001rs}. Substituting $\sigma^{0}_{NP}\left(s\right)$ in Eq.\,\ref{Eq:eenunuA} while appropriately changing the variables and integrating over the convolution function $H$, one can evaluate the full cross section for the new physics interactions as well as for the SM piece. To compare our result with the experimentally measured data we first add the NP and SM contribution together to obtain the total cross section for the underlying process \emph{i.e.} $\sigma(s) =  \sigma_{NP}(s) + \sigma_{SM}(s)$. To derive the constraints from the experimental results, one can interpret $|\sigma-\sigma_{exp}| $ $\leq$ $ \delta \sigma_{exp} $, where $ \sigma_{exp} $ $\pm$ $ \delta \sigma_{exp}$ is the experimental measured cross section for the process $e e \to \nu \bar{\nu} \gamma$. Expanding the $\sigma$ with the NP and SM contributions and dividing both side with $\sigma_{SM}$ one can write \cite{Berezhiani:2001rs}, 
\begin{eqnarray}
	\bigg|1+\frac{\sigma_{NP}}{\sigma_{SM}}-\frac{\sigma_{exp}}{\sigma_{SM}}\bigg| \leq \bigg(\frac{\sigma_{exp}}{\sigma_{SM}}\bigg) \bigg(\frac{\delta \sigma_{exp}}{\sigma_{exp}}\bigg)
\end{eqnarray}
\noindent
For simplification purpose, if we further assume the central value of the experimental measurement is same as $\sigma_{SM}$, then the above equation can be written in the following fashion
\begin{equation}
	\bigg|\frac{\sigma_{NP}}{\sigma_{SM}}\bigg| \leq  \bigg(\frac{\delta \sigma_{exp}}{\sigma_{exp}}\bigg)
	\label{Eq:Monoexp}
\end{equation}       
From Eq.\,\ref{Eq:Monoexp}, one can translate this inequality to impose bound in the $M_{\chi}$ vs $\mathcal{Y}_{c}$ plane. To do so, we adopt the results from \cite{L3:2003yon}, and considering the veto $14^{\circ} < \theta_{\gamma} < 166^{\circ}$, $E_{\gamma} > 1 \text{GeV}$ and $p^{\gamma}_{T} > 0.02s$. The measured experimental value we have used is, $\sigma_{exp}(\text{pb}) = 4.29\pm0.85\pm0.05$ for $\sqrt{s} = 208$ GeV, where the first uncertainty on $\sigma_{exp}$ is statistical and the second is systematic.  The exclusion region for this bound is illustrated as the yellow shaded region in Fig.\,\ref{Fig:collider_bound}.      

\subsubsection{BBN constraints}
The neutrinos in this model obtain the required eV scale Dirac mass via loop-mediated process. The $\nu_R$ is the right chiral component of the Dirac field. In principle, this new degrees of freedom, $\nu_{R}$, can populate the early universe via $\ell^{+} \ell^{-} \leftrightarrow \nu_{R} \bar{\nu_{R}}$, with cross-section denoted as $\sigma_R$, where the process is mediated through t-channel $\chi^{\pm}$ exchange. The Big Bang Nucleosynthesis (BBN) suggests the new relativistic degrees of freedom must lie within the range $\Delta N_{eff} \equiv N_{eff}-3.046 = 0.10^{+0.44}_{-0.43}$ at 95\% confidence level with the combination of He + Planck TT +  lowP + BAO Dataset \cite{Planck:2015fie}. Using \cite{Planck:2015fie,Dey:2018yht,Davidson:2009ha}, one can recast this limit for our model and impose constraints in the $M_{\chi} - \mathcal{Y}_{c}$ plane. To satisfy this condition, we demand the right-handed neutrinos decouple from the thermal bath before the quark-hadron transition. Setting the decoupling temperatures for right handed neutrinos is $T_{d,\nu_{R}}\approx200~\text{MeV}$ and for left handed neutrinos is $T_{d,\nu_{L}}\approx 3~\text{MeV}$ \cite{Olive:1999ij} respectively, one can put an upper bound on $\ell^{+} \ell^{-} \to \nu_{R} \bar{\nu}_{R}$ cross section via \cite{Steigman:1979xp} following relation, 
\begin{equation}
	(T_{d,\nu_{R}}/T_{d,\nu_{L}})^3\approx(\sigma_L/\sigma_R) = (2 M_\chi^\pm/v_L \mathcal{Y}^{i}_c |V_{\ell i}|)^4
\end{equation}
where $|V_{\ell i}|$ is the right-handed counterpart of the $U_{PMNS}$ matrix which is required to diagonalise the neutrino mass matrix (see Eq.\,\ref{Eq:Nu_diag}) and $v_{L}$ is the \emph{vev} that invoke the symmetry breaking of $SU(2)_{L}$ gauge group which is 246 GeV. Fixing $|V_{\ell i}|$ at a moderate value 0.5, we deduce the corresponding bound and illustrate the exclusion region in the Fig.\,\ref{Fig:collider_bound} as a pink shading.

\subsection{Charged Higgs Boson Decay and Production}
With a large number of scalars: neutral, singly and doubly charged, the model has a rich phenomenological implication. Though the presence of doubly charged scalar makes the model more interesting, its interaction with the SM leptons are proportional to the heavy-light lepton mixing, and is not quite promising. However the singly charged scalar ($ \chi^\pm $) has direct  interaction with the SM leptons and is independent of the heavy-light lepton mixing. Thus we have focused mainly on this singly charged scalar. We considered the following  mass hierarchy of the singly charged scalars: $ M^{2}_{\chi^\pm} <M^{2}_{\zeta^\pm_{L}} < M^{2}_{\zeta^\pm_{R}}$. The possible partial decay widths of $ \chi^\pm $ is given in Eq.\,\ref{Eq:chidecaywidth}. 
\begin{eqnarray}
\Gamma\left(\chi^{-}_{m} \to \ell^{-}_{mp}\nu_{mk} \right) & =& \frac{1}{8\pi M^{3}_{\chi}}\left|2~\mathcal{Y}^{ij}_{c}U_{PMNS}^{ik}U^{L~jp}_{e_{11}}\right|^{2}\left(M^{2}_{\chi} - m^{2}_{\ell_p}\right)\lambda^{\frac{1}{2}}\left(M^{2}_{\chi}, m^{2}_{\ell_p}, m^{2}_{\nu_k} \right) \nonumber \\
\Gamma\left(\chi^{-}_{m} \to \bar{u}_{mk}d_{mp} \right) & = & \frac{3}{8\pi M^{3}_{\chi}}\left|\mathcal{Y}^{ij}_{q}U^{L*~ik}_{u_{21}}U^{R~jp}_{d_{21}}\right|^{2}\left(M^{2}_{\chi} - m^{2}_{u_k} - m^{2}_{d_p}\right) \nonumber \\&& \lambda^{\frac{1}{2}}\left(M^{2}_{\chi}, m^{2}_{u_k}, m^{2}_{d_p} \right) 
\label{Eq:chidecaywidth}
\end{eqnarray}
From the aforementioned, Eq.\,\ref{Eq:chidecaywidth}, it can be seen that the singly charged scalar, $\chi^\pm$, has two possible decay modes: $l\,\bar{\nu}$ and $\bar{u}\,d$, where the flavor indices are suppressed. While for the former, the light-light lepton mixing $U^L_{e_{11}}$ enters in the expression of partial decay width, for the later decay mode both $U^L_u$ and $U^R_d$ enters in the expression. The Yukawa coupling, $\mathcal{Y}_{q}$, which is a free parameter of the model can be chosen of the $\mathcal{O}[10^{-2}]$ such that the leptonic mode become the most dominant decay channel.

The analytical expression for the $\chi^\pm$$\chi^\mp$ pair production at $e^+e^-$ collider is given by \cite{Djouadi:1996pj,Komamiya:1988rs},
\begin{eqnarray}
	\sigma_{e^+e^- \rightarrow \chi^\pm\chi^\mp} = 
	\frac{2 G_F^2M_W^4 s^4_{\theta_{W}}}{3\pi s}\Bigg[1+\frac{4s^2_{\theta_{W}}-1}{2c^2_{\theta_{W}}\big(1-\frac{M_Z^2}{s}\big)}+\frac{8s^4_{\theta_{W}}-4s^2_{\theta_{W}}+1}{8c^4_{\theta_{W}}\big(1-\frac{M_Z^2}{s}\big)^2}\Bigg]\Big(1-\frac{4 M^2_{\chi^\pm}}{s}\Big)^{3/2}
\end{eqnarray}

\begin{figure}[h]
	\centering
	\includegraphics[height=10cm,width=10cm]{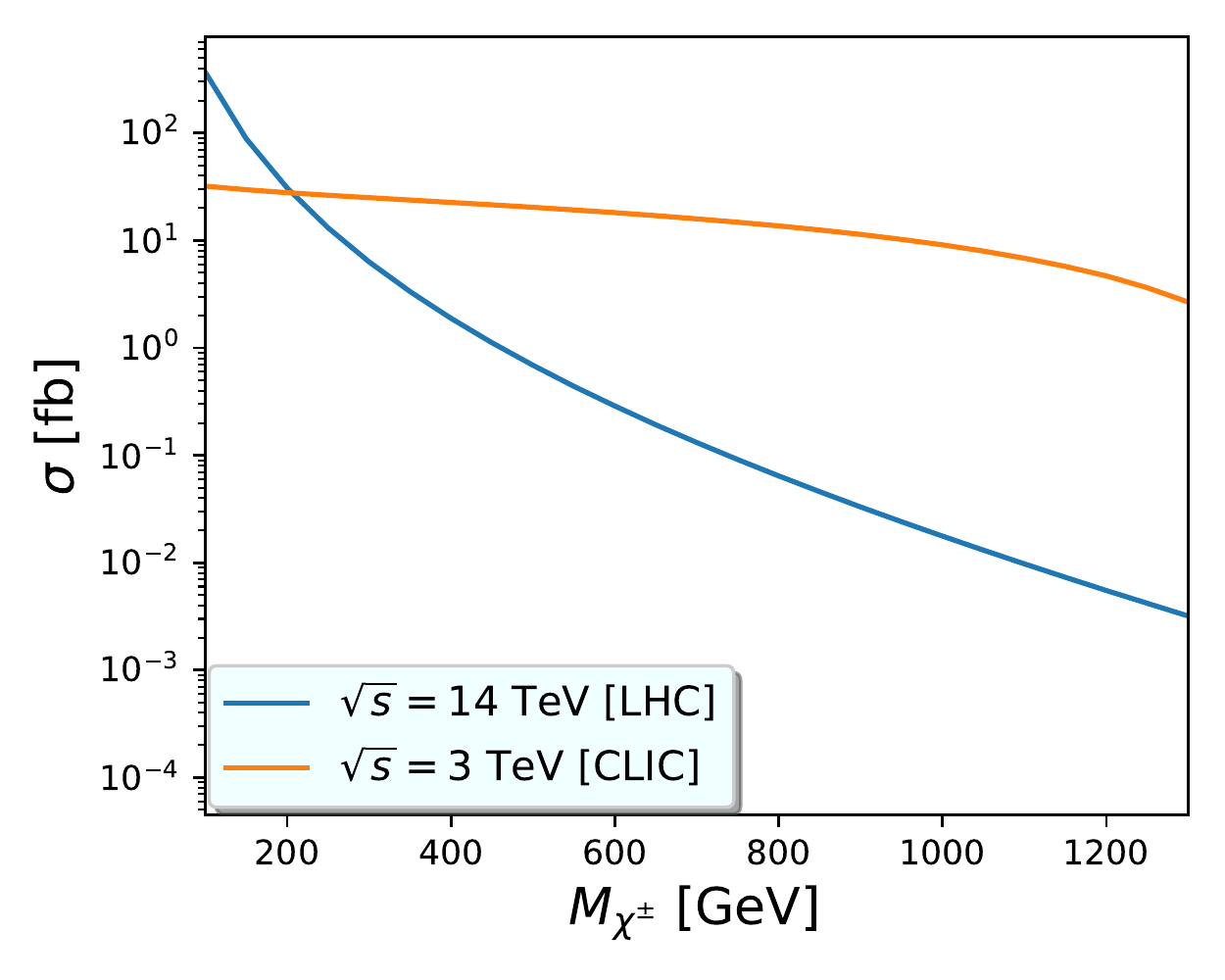}
	\caption{Production cross-section of $\chi^{\pm}$ at $\sqrt{s}$ = 14 TeV LHC and $\sqrt{s}$ = 3 TeV CLIC. }
	\label{Fig:prodcrx_chi}
\end{figure}

  We have implemented the model file in \texttt{FeynRules} \cite{Alloul:2013bka} and used \texttt{MadGraph} \cite{Alwall:2014hca} to get the leading order cross sections. For comparison purpose, we have presented the cross-section for pair-production of $\chi^\pm$ at 14 TeV LHC and at 3 TeV CLIC in Fig.\,\ref{Fig:prodcrx_chi}. For $ M_{\chi^\pm} = 500$ \text{GeV} at 14 TeV LHC and at 3 TeV CLIC the cross sections are obtained to be 0.69 fb and 20.3 fb respectively. Again from the figure it is clear that with increase in the mass of $\chi^\pm$   the cross section for LHC falls more sharply as compared to CLIC.

\subsubsection{Associated production of $\chi^\pm$ from decay}
The model also contains many heavy vector like fermions, important for the mass generation of the SM fermions. For our choice of masses, a pair of the heavy vector like top and its subsequent decay can also produce charged scalar $\chi^\pm$ hence increasing the total production cross-section of $\chi^\pm$. The partial decay widths of the $heavy-top$ T and $heavy-bottom$ B are expressed as follows,
\begin{eqnarray}
	\Gamma(T \rightarrow t h_{sm}) &=& \frac{1}{16 \pi^2 M_T^3} \frac{1}{4} |\mathcal{Y}_u|^2 (M_T^2 + m_t^2 - m_{h_{sm}}^2) \lambda^{\frac{1}{2}}\left(M_T^2, m_t^2 , m_{h_{sm}}^2 \right) \nonumber \\
	\Gamma(T \rightarrow t H^0) &=&  \frac{1}{16 \pi^2 M_T^3} \frac{1}{4} |\mathcal{Y}_u|^2 (M_T^2 + m_t^2 - m_H^2) \lambda^{\frac{1}{2}}\left(M_T^2, m_t^2 , m_H^2 \right)\nonumber \\
	\Gamma(T \rightarrow b W^{\pm}) &=& \frac{G_F M_T^2}{8 \pi \sqrt{2}} |V_{tb}|^2 |U_{Tt}|^2 \left(1 - \frac{M_W^2}{M_T^2}\right)^2 \left(1 + 2 \frac{M_W^2}{M_T^2}\right) \nonumber \\ 
	\Gamma(T \rightarrow B \chi^{\pm}) &=&\frac{1}{16 \pi^2 M_T^3} |\mathcal{Y}_{q}|^2 ((M_T + M_B)^2 - M_\chi^2) \lambda^\frac{1}{2}(M_T^2, M_B^2, 
	m_\chi^2) \nonumber \\ 
		\Gamma(B \rightarrow b h_{sm}) &=& \frac{1}{16 \pi^2 M_B^3} \frac{1}{4} |\mathcal{Y}_d|^2 (M_B^2 + m_b^2 - m_{h_{sm}}^2) \lambda^{\frac{1}{2}}\left(M_B^2, m_b^2 , m_{h_{sm}}^2 \right) \nonumber \\
	\Gamma(B \rightarrow b H^0) &=&  \frac{1}{16 \pi^2 M_B^3} \frac{1}{4} |\mathcal{Y}_d|^2 (M_B^2 + m_b^2 - m_H^2) \lambda^{\frac{1}{2}}\left(M_B^2, m_b^2 , m_H^2 \right)\nonumber \\
	\Gamma(B \rightarrow t W^{\pm}) &=& \frac{G_F M_B^2}{8 \pi \sqrt{2}} |V_{tb}|^2 |U_{Bb}|^2 \left(1 - \frac{M_W^2}{M_B^2}\right)^2 \left(1 + 2 \frac{M_W^2}{M_B^2}\right)
\end{eqnarray}
 
\begin{figure}[H]
	\centering
	\includegraphics[height=6.5cm,width=8cm]{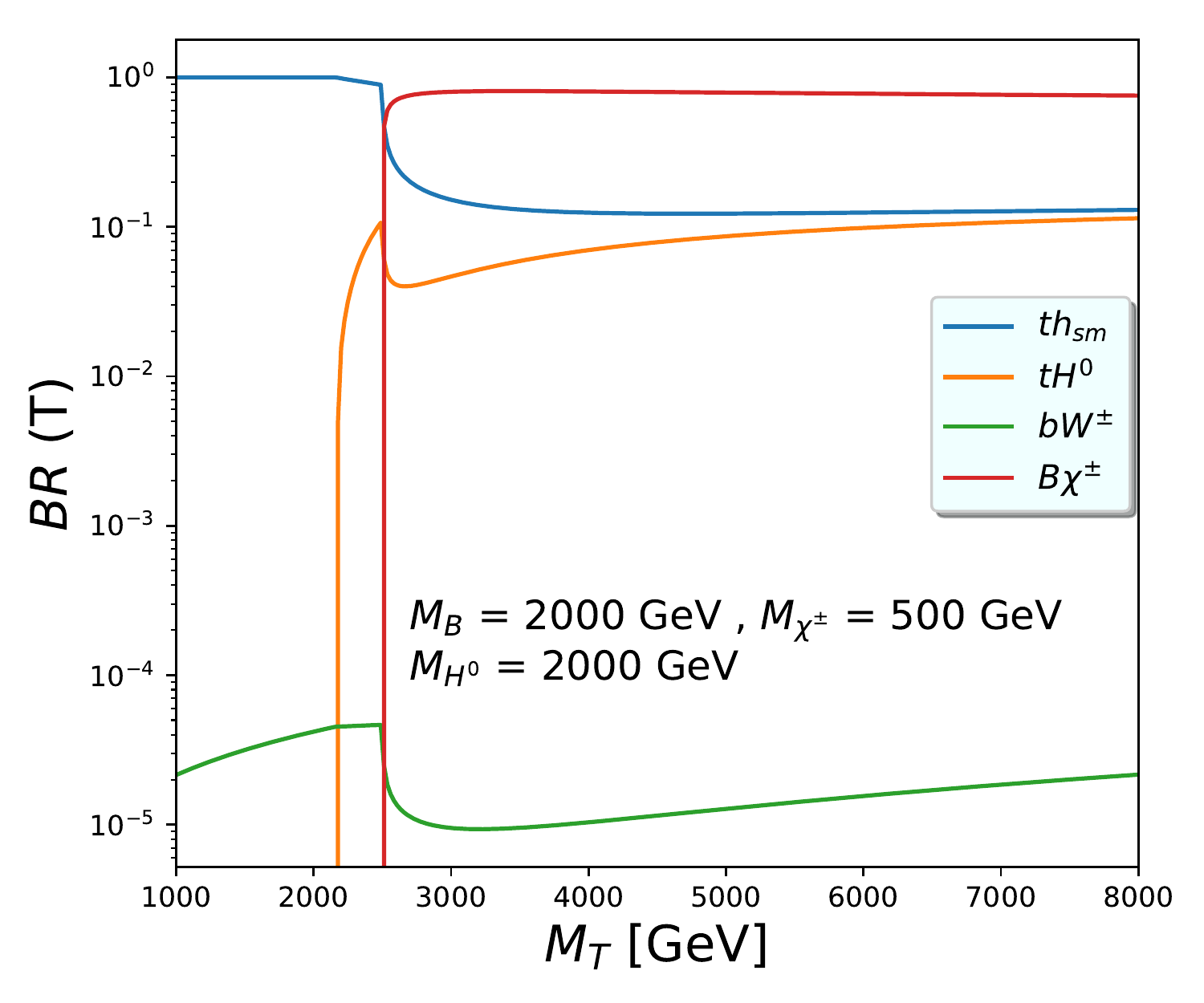}	
	\includegraphics[height=6.5cm,width=8cm]{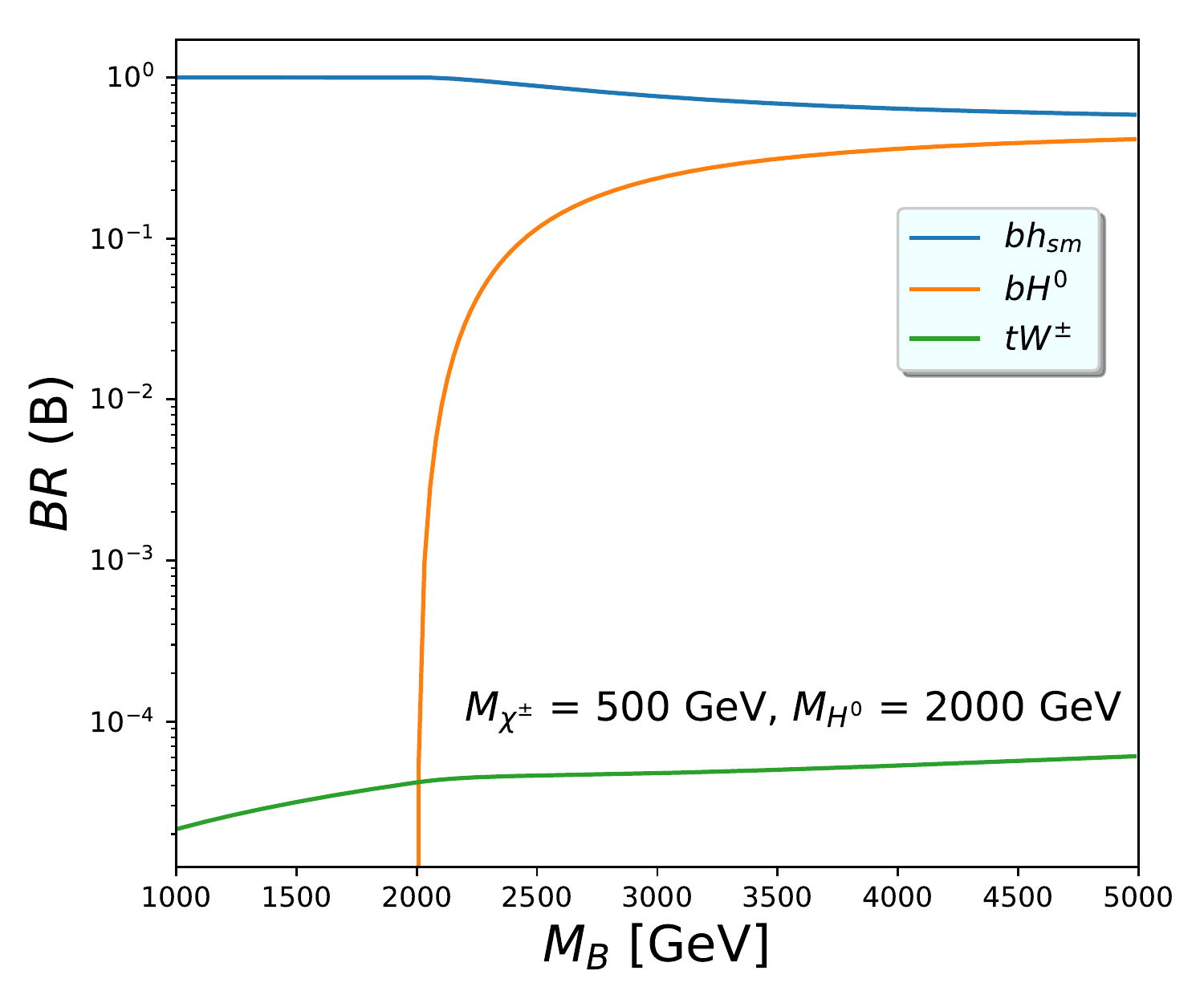}		\caption{Branching fractions  of $heavy-top$ quark (T) and $heavy-bottom$ quark (B).  }
	\label{Fig:BR_heavytop}
\end{figure}

In the left panel of Fig.\,\ref{Fig:BR_heavytop}~we show the branching fractions of $heavy-top$ T, considering the $heavy-bottom$ mass $M_B$ = 2 TeV, $M_\chi^\pm$ = 500 GeV and heavy neutral scalar mass $M_{H^0}$ = 2 TeV respectively. In the right panel of Fig.\,\ref{Fig:BR_heavytop}~ we have shown the branching fractions of $heavy-bottom$ B. The Yukawa couplings, $\mathcal{Y}_{q}$ and $ \mathcal{Y}_{u,d}$, are set to 1 and 0.01 respectively.

From Fig.\,\ref{Fig:BR_heavytop}, its clear that before the threshold of $heavy-bottom$ B and the heavy neutral scalar $H^0$ the dominant decay mode of the $heavy-top$ T will be the $t\,h_{sm}$ mode. After the $heavy-bottom$ B threshold is reached the dominant decay model of the $heavy-top$ will be the $\text{B} \chi^\pm$ mode. Similarly for before the heavy neutral scalar $H^0$ threshold the $heavy-bottom$ will decay dominantly into $b\, h_{sm}$ mode. 

\begin{figure}[H]
	\centering
	\includegraphics[height=8cm,width=10cm]{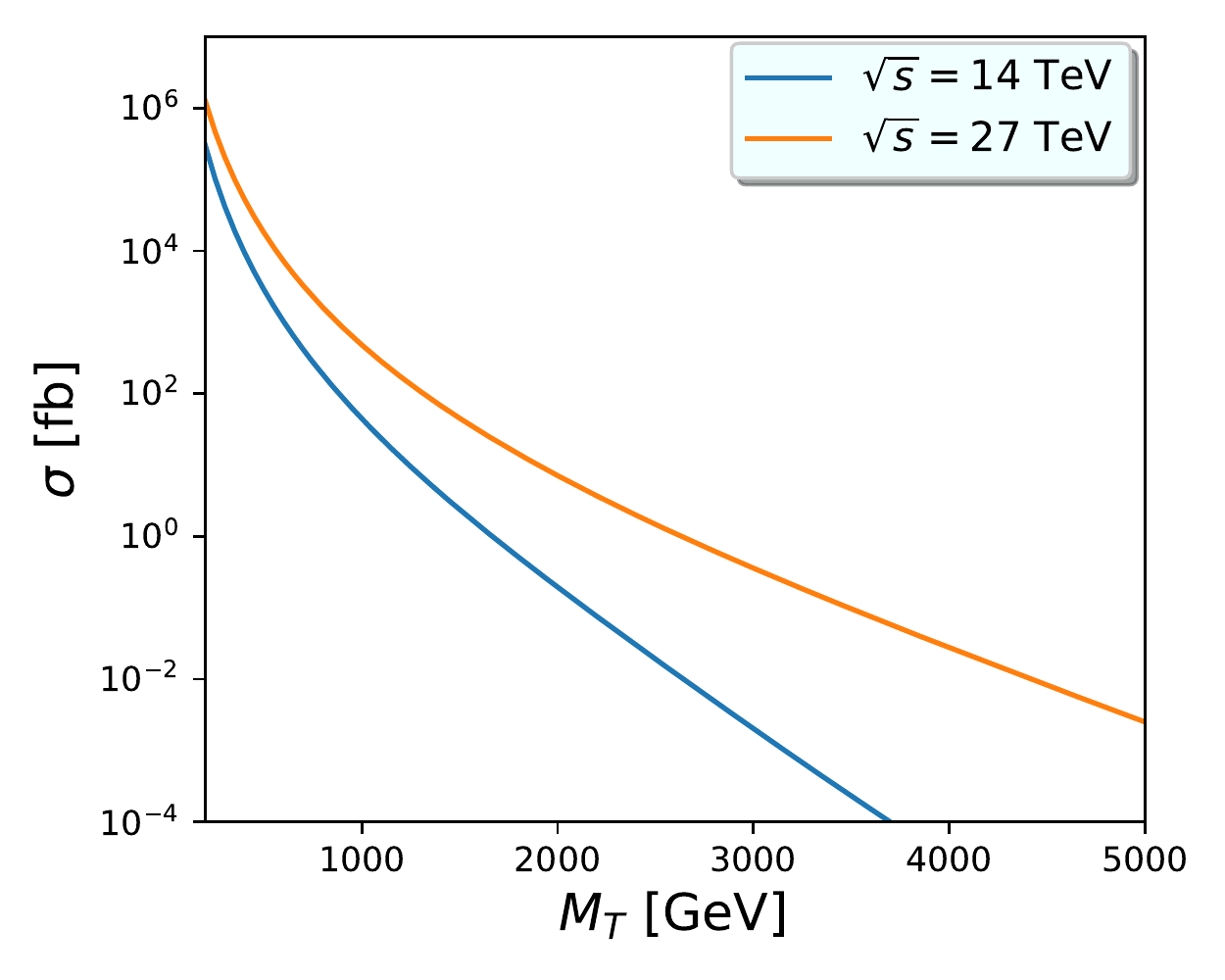}	
	\caption{ Production cross-section of $heavy-top$ quark (T) for $\sqrt{s}$ = 14, 27 TeV LHC respectively.  }
	\label{Fig:prodcrx_heavytop}
\end{figure}

Fig.\,\ref{Fig:prodcrx_heavytop}~ shows the T$\bar{\text{T}}$ pair-production cross-section at 14  LHC, and at a $pp$ machine with $\sqrt{s} = 27$ TeV respectively. For higher mass of the $heavy-top$ quark (T), a larger center of energy is more suitable, which is quite apparent from the figure.

With the information of the production cross section of the $heavy-top$ pair and their respective decay modes, we predict the most efficient signal, a multi b-jet as well as multi lepton final state, for the $heavy-top$ T discovery at LHC.  In Table.\,\ref{Tab:LHC_prediction}, we have shown the predicted cross section for $6b$ + $2l$ + $\slashed{E}_T$ final state resulting from the consequent decay of the produced $heavy-top$ T at 14 and 27 TeV LHC respectively. We have considered $M_T = 3 \,\text{TeV}$, $M_B = 2 \,\text{TeV}$, $M_{\chi^\pm} = 500 \,\text{GeV}$.

 \FloatBarrier
\begin{table}[H]
	\centering
	\resizebox{0.5\textwidth}{!}{
		\begin{tabular}{|c|c|c|}
			\hline
			$\sqrt{s}$ & final state & cross-section \\
			\hline
			\hline
			14 TeV&
			\pbox{10cm}{6b + $2l$ + $\slashed{E}_T$} & 0.72828$\times10^{-3}$ fb\\
			\hline
			\hline
			27 TeV&
			\pbox{10cm}{6b + $2l$ + $\slashed{E}_T$} & $0.129132$ fb\\
			\hline
		\end{tabular}
	}
	\caption{In the above table, we have shown the predicted cross section for 6b + $2l$ + $\slashed{E}_T$ final state resulting from the consequent decay of the pair produced $heavy-top$ T at 14 and 27 TeV LHC respectively. Below the kinematic threshold the produced $heavy-top$ T decays to $\text{B} \chi^\pm$ and the $heavy-bottom$ B then decays to $b\, h_{sm}$. Again the produced $\chi^\pm$ decays to $l+\slashed{E}_T$ final state, leading to 6b + $2l$ + $\slashed{E}_T$ signal resulting from the pair produced $heavy-top$ T at LHC. The above cross sections are calculated by considering the following masses: $M_T$ = 3\,TeV, $M_B$ = 2\,TeV, $M_{\chi^\pm}$ = 500\,GeV.}
	\label{Tab:LHC_prediction}
\end{table}

\section{Summary}\label{Sec:conclu}
We have discussed an alternate variant of Left-Right symmetric model, embedding Dirac type neutrinos, where the small neutrino masses have been generated radiatively. In the absence of any bi-doublet scalar charged under both $SU(2)_L$ and $SU(2)_R$, we consider a universal seesaw  scheme for the mass generation of  SM like charged fermions; the realization of such a seesaw mechanism is performed by heavy vector-like  fermions. In the absence of any gauge singlet neutrino, light neutrinos acquire their masses radiatively. For  neutrino mass generation at one-loop, we require additional scalar multiplets. We consider $3\sigma$ variation of neutrino oscillation data and find out  the relevant parameter space, in which neutrino mass and mixings are satisfied. We find that 
	\begin{itemize}
		\item
		With the increase in  heavier  gauge singlet charged scalar's mass $M_{\chi^{\pm}}$, diagonal
		Yukawa couplings $\mathcal{Y}_{z_{22,33}}$  decrease in order to satisfy neutrino oscillation data.  
		\item
		For both normal and inverted mass hierarchy, there are ample parameter space, where neutrino oscillation data is satisfied. 
	\end{itemize}
	We furthermore perform a detailed analysis of the scalar sector, mostly focussing on the lightest charged scalar $\chi^{\pm}$. In our analysis, we consider different direct as well as indirect search constraints applicable on $\chi^{\pm}$ including ATLAS di-lepton+MET search \cite{ATLAS:2019lff}, LEP mono-photon search \cite{L3:2003yon}, and the constraints from Big-Bang Nucleosynthesis \cite{Planck:2015fie}. We find that the BBN constraint can rule out a significant parameter space, 
	\begin{itemize} 
		\item
		In particular,  Yukawa coupling $\mathcal{Y}_c$, which is the coupling between $\chi^{\pm}$ and a lepton and neutrino,  larger than 0.3 (0.7) are ruled out for $M_{\chi^{\pm}} > 400 (1000)$ GeV from BBN. The LEP mono-photon constraint is relatively relaxed than the BBN bounds. 
		\item
		ATLAS di-lepton+MET search rules out $M_{\chi^{\pm}} < 320 $ GeV independent of the parameter $\mathcal{Y}_c$. 
		\item
		The LFV in our case does not impose any serious constraint on the model parameters. 
	\end{itemize}
	The extended scalar sector which is required for the neutrino mass generation at one-loop via Fig.\,\ref{Fig:NeutrinoG},\, \ref{Fig:NeutrinoM}, has rich phenomenological significance. Out of the three singly charged scalars ($\zeta^{\pm}_{L},\zeta^{\pm}_{R},\chi^\pm$),  the singlet one $\chi^\pm$ has direct interaction with two SM leptons and hence can give rise to a $l+\slashed{E}_T$ and $qq^{\prime}$ final states. We discussed its collider phenomenology such as production,  decay and branching ratios  in detail. We find that 
	\begin{itemize}
		\item
		For $M_{\chi^{\pm}}$ between 250-1000 GeV, the pair-production cross-section at 14 TeV LHC varies as $\sigma \sim 10^2-10^{-1}$ fb. For a  $e^+e^-$ machine with c.m.energy 3 TeV, cross-section is $\sigma \sim 10$ fb.
	\end{itemize}
	Furthermore, we also consider the associated production of $\chi^{\pm}$ from the decay of a  heavy vector like top-quark T. We find that, the heavy vector-like $T$ quark of mass $\sim 1$ TeV can copiously be produced at the HL-LHC with $\sigma \sim 10^2 $ fb, and decay pre-dominantly to a $\chi^{\pm}$ and heavy vector-like $B$ state. A more sophisticated collider analysis of the BSM particles will be our future goal.

\section{Acknowledgment}\label{Sec:Acknowledge}
The authors acknowledge the support from the Indo-French Centre for the Promotion of Advanced Research (Grant no: 6304-2).

\appendix
\section{\label{App:One_loop}One-loop Calculation of Neutrino Mass}
From Fig.\,\ref{Fig:NeutrinoM}, the loop integral with the SM charged leptons mediating the loop will be \cite{Bouchand:2012dx},
\begin{eqnarray}
-i\Sigma_{mn} & =&  \int\frac{d^{4}k_1}{\left(2\pi\right)^{4}}\left( \frac{i}{\gamma^{\mu}k_{\mu1} - m_{e_i}}\right) \frac{i}{(p-k_1)^2 - m^2_{h_{j}}} U^{h^\pm}_{1j}U^{h^\pm *}_{3j}\xi_{mn} \nonumber \\
		      & =& \int\frac{d^{4}k_{1}}{\left(2\pi\right)^{4}}\xi_{mn}U^{h^\pm}_{1j}U^{h^\pm *}_{3j}\frac{m_{e_i}}{\left(k^{2}_{1} - m^{2}_{e_i}\right)\left(k^{2}_{1} - m^{2}_{h_j}\right)} \nonumber \\
		      & =& \xi_{mn}U^{h^\pm}_{1j}U^{h^\pm *}_{3j}\frac{i}{16\pi^{2}}\mathcal{B}_{0}\left(p^2 = 0, m^{2}_{e_i}, m^{2}_{h_j}\right) \nonumber \\
		      & =& \xi_{mn}U^{h^\pm}_{1j}U^{h^\pm *}_{3j}\frac{i}{16\pi^{2}}\left[\Delta + 1 + \ln\left(\frac{m^{2}_{e_i}}{\mu^{2}}\right) + \frac{m^{2}_{h_j}}{m^{2}_{h_j} - m^{2}_{e_i}}\ln\left(\frac{m^{2}_{h_j}}{m^{2}_{e_i}}\right) \right]
\end{eqnarray}

\noindent
where
\begin{itemize}
\item[] $\Delta = \frac{2}{\epsilon} - \ln\pi + \gamma_{E} \footnote{For details please see Ref.\,\cite{Passarino:1978jh,Weinzierl:2006qs}.}$
\item[] $\mu =$ Scale of renormalisation
\item[] $\xi_{mn} = \mathcal{Y}^{mb}_{c}U^{L}_{11b\alpha}U^{R*}_{21\beta \alpha}\mathcal{Y}^{n\beta}_{z}$
\item[] $\mathcal{B}_{0}$ is the \textit{Passarino-Veltman} function and its expression can be found in the literature \cite{tHooft:1978jhc}
\end{itemize}
\noindent
with further simplification one can obtain the mass contribution coming from $m_{\nu}\left(e_{i}\right)^{mn}$ term. Similarly the rest of the three contribution can be calculated.

\bibliographystyle{utphys}
\bibliography{bibitem}
\end{document}